\documentclass[journal]{IEEEtran}

\usepackage{titlesec}
\titlespacing\subsection{0pt}{3pt plus 4pt minus 2pt}{4pt plus 2pt minus 2pt}
\usepackage{xcolor,soul,framed}
\colorlet{shadecolor}{yellow}

\usepackage[pdftex]{graphicx}
\graphicspath{{../pdf/}{../jpeg/}}
\DeclareGraphicsExtensions{.pdf,.jpeg,.png}

\usepackage{tabularx}
\usepackage[font=small,skip=0ex]{caption}
\usepackage{array}
\usepackage{amsfonts}
\usepackage{amsmath}
\usepackage{cases}
\usepackage[ruled,linesnumbered]{algorithm2e}
\usepackage{nicematrix}
\usepackage[noadjust]{cite}
\usepackage{bm}
\usepackage{subfig}

\begin{document}
\bstctlcite{IEEEexample:BSTcontrol}

\title{Throughput Characterization of Wireless CSMA Networks With Arbitrary Sensing and Interference Topologies}

\author{
Xinghua Sun,~\IEEEmembership{Member,~IEEE,}
Wenhai Lin,~\IEEEmembership{Graduate Student Member,~IEEE,}
Ruike Zhou
\thanks{Xinghua Sun, Wenhai Lin and Ruike Zhou are with the School of Electronics and Communication Engineering, Sun Yat-sen University, Shenzhen 518107, China (e-mail: sunxinghua@mail.sysu.edu.cn; linwh33@mail2.sysu.edu.cn; zhourk3@mail2.sysu.edu.cn).}
}

\maketitle
\begingroup
\renewcommand\thefootnote{}
\footnotetext{This work has been submitted to the IEEE for possible publication. Copyright may be transferred without notice, after which this version may no longer be accessible.}
\endgroup

\begin{abstract}
The performance analysis of wireless CSMA networks is notoriously difficult due to the intricate sensing and interference relationships among links. Even the fundamental problem of throughput characterization remains open when sensing and interference topologies are both arbitrary. In this paper, we develop a new analytical framework for throughput characterization in wireless CSMA networks with arbitrary sensing and interference topologies. The proposed framework yields explicit throughput expressions without relying on the commonly adopted zero-propagation-delay assumption. The key idea is to exploit the clique structure of the sensing graph to transform the original CSMA network into an equivalent multi-channel network, and then model its dynamics through a discrete-time Markov renewal process. In this way, the framework explicitly captures global coupling among links and enables analytical evaluation of how access parameters affect network performance. The proposed analysis is applied to several representative CSMA scenarios, including networks with multi-BSS IEEE 802.11 networks with universal frequency reuse, and ad-hoc topologies exhibiting hidden-terminal, exposed-terminal, and flow-in-the-middle effects. Simulation results show that, in dense deployments and in scenarios with strong coupling among link behaviors, the proposed model significantly outperforms existing analytical approaches in throughput estimation and enables more accurate determination of access parameters.
\end{abstract}

\begin{IEEEkeywords}
Carrier Sense Multiple Access~(CSMA), wireless networks, throughput analysis, hidden terminals, multi-BSS IEEE 802.11 networks, Markov renewal process.
\end{IEEEkeywords}

\IEEEpeerreviewmaketitle

\section{Introduction}
\label{sect:intro}

\IEEEPARstart{C}{arrier} Sense Multiple Access (CSMA) is a fundamental random-access mechanism that has been widely adopted in modern wireless networks. Under CSMA, a transmitter senses the channel before initiating transmission and defers its access if an ongoing transmission is detected~\cite{kleinrock1975packet}. A wireless CSMA network typically consists of multiple transmitter--receiver pairs, each forming a communication link. Such networks arise in a wide range of practical settings. For example, in Wi-Fi networks, multiple stations contend for access to a common access point, whereas in ad-hoc networks, each station may communicate with a different receiver.

Despite its practical success, the performance analysis of wireless CSMA networks remains notoriously difficult. The main challenge lies in the intricate sensing and interference relationships among links. In particular, due to the limited carrier-sensing range, not all transmitters can detect one another. As a result, each transmitter observes only a local subset of network activity, and the medium-access behavior of different links becomes strongly coupled.

\begin{figure}[t]
	\centering
    \includegraphics[width=0.35\textwidth]{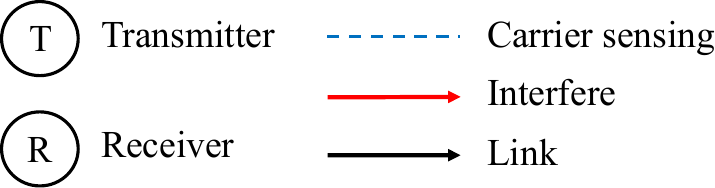}
    \hfill
    \subfloat[]{
		\includegraphics[height=0.14\textwidth]{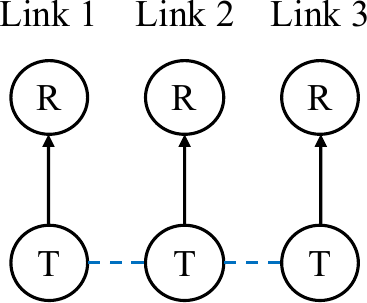}
        \label{fig:topo_flow_midd}}
    \hfill
	\subfloat[]{
		\includegraphics[height=0.14\textwidth]{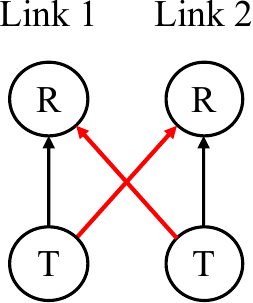}
        \label{fig:topo_hide_term}}
    \hfill
    \subfloat[]{
		\includegraphics[height=0.14\textwidth]{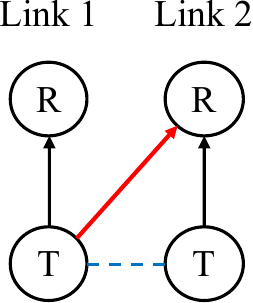}
        \label{fig:topo_expo_term}}
	\caption{Representative topological scenarios in wireless CSMA networks: (a) flow-in-the-middle, (b) hidden-terminal, and (c) exposed-terminal.}
	\label{fig:first_topo}
\end{figure}

A representative example is the \emph{flow-in-the-middle} problem shown in Fig.~\ref{fig:topo_flow_midd}. In this topology, link~2 is the \emph{flow-in-the-middle} link: its transmitter is within the carrier-sensing range of the transmitters of both links~1 and~3, whereas links~1 and~3 cannot sense each other directly. Consequently, link~2 must defer whenever either of the other two links transmits, while links~1 and~3 may still access the channel independently. This asymmetric contention relationship puts link~2 at a disadvantage and may lead to severe throughput starvation. The example also shows that even links without direct sensing relationships may still be statistically coupled through the network topology.

The analysis becomes even more involved because sensing relationships generally do not coincide with interference relationships. Since the transmitter and receiver of a link are geographically separated, whether two links can sense each other does not necessarily determine whether they interfere with each other. This mismatch gives rise to the well-known hidden-terminal and exposed-terminal problems. As shown in Fig.~\ref{fig:topo_hide_term}, the transmitters of links~1 and~2 cannot sense each other, yet their simultaneous transmissions collide at the receivers, resulting in a hidden-terminal problem. In contrast, Fig.~\ref{fig:topo_expo_term} illustrates an exposed-terminal scenario, in which the transmitters of links~1 and~2 can sense each other, although the transmission of link~2 does not interfere with the receiver of link~1.

Under such topologies, even when the classical collision model is adopted, i.e., a transmission succeeds if and only if no interfering link transmits concurrently, throughput characterization remains highly nontrivial. The success probability of a link depends on the joint activity of multiple other links. Because these links may be indirectly coupled through carrier sensing, the probability of simultaneous transmissions cannot, in general, be factorized into independent marginal probabilities. Therefore, accurate throughput analysis requires a model capable of capturing such \emph{global coupling} across the network.

These challenges are common in CSMA-based ad-hoc networks, where nodes are geographically distributed and only partially aware of ongoing transmissions. Similar complexity also arises in infrastructure networks. In dense multi-BSS WLANs, for example, different stations may observe different subsets of ongoing transmissions, and their access behavior may be strongly correlated through overlapping sensing domains. In such scenarios, the mismatch between carrier sensing and actual interference becomes increasingly pronounced, rendering the analytical characterization of practical medium-access behavior much more challenging.

A fundamental and still open problem is therefore how to characterize the throughput of CSMA networks under arbitrary sensing and interference relationships. In this paper, we develop a new analytical model for wireless CSMA networks to address this problem. 
The key idea of the proposed analysis is to represent the network topology using a sensing graph and an interference graph, and then transform the original single-channel CSMA network into an equivalent multi-channel network by exploiting the clique structure of the sensing graph\footnote{The terms single-channel and multi-channel refer to whether the network has one or multiple available channels for nodes to access. In this paper, all nodes operate on the same frequency band; hence, the original CSMA network is single-channel.}. Based on this transformation, we establish a discrete-time Markov renewal process to model the activity of the equivalent multi-channel network. Since the model directly tracks network-wide state evolution, it can explicitly capture global coupling among links. From the steady-state probabilities of network states, we further derive explicit expressions for link throughput and use them to analyze the impact of access parameters on network performance.

Building upon the proposed framework, we examine several representative CSMA scenarios, including multi-BSS IEEE 802.11 networks and ad-hoc topologies exhibiting hidden-terminal, exposed-terminal, and flow-in-the-middle effects. These case studies demonstrate how the proposed framework can be applied to practical wireless networks with strong spatial coupling and sensing--interference mismatch. We also analyze the computational complexity of the proposed framework and discuss techniques to enhance tractability.

The remainder of this paper is organized as follows. Section~II reviews related work on the modeling of wireless CSMA networks. Section~III introduces the system model. Section~IV presents the proposed analytical framework. Section~V derives the link-throughput expressions. Section~VI provides case studies and simulation results. Section~VII discusses the computational complexity of the proposed model. Finally, Section~VIII concludes the paper.

\begin{figure*}[t]
  \centering
  \includegraphics[width=0.7\textwidth]{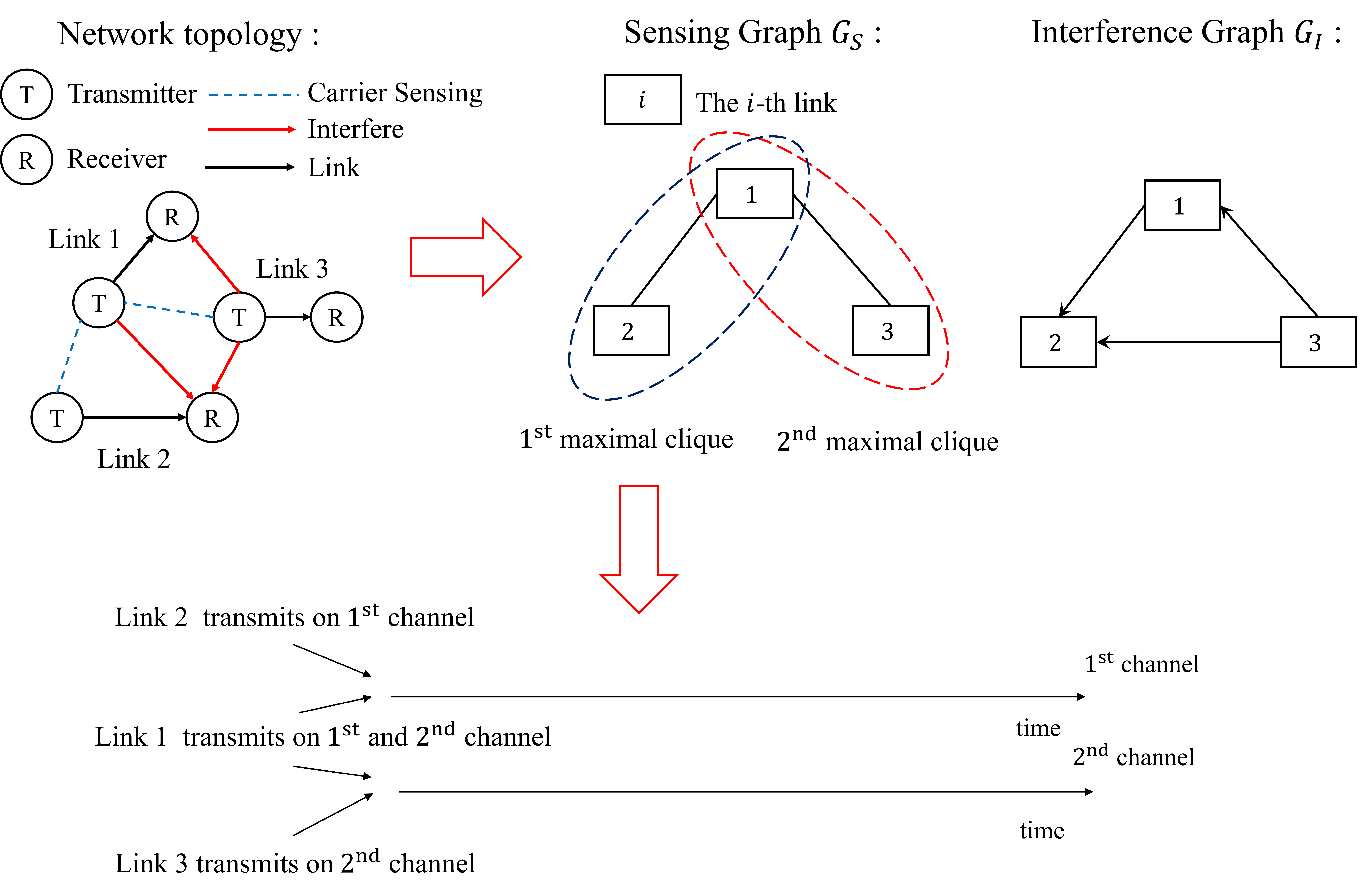}
  \caption{Illustration of the sensing graph \(G_S\), the interference graph \(G_I\), and the equivalent multi-channel network.}
   \label{fig:network_topology}
\end{figure*}

\section{Related Work}
\label{sect:related}

Early studies on wireless CSMA networks mainly focused on multi-access scenarios~\cite{kleinrock1975packet,bianchi2000performance,malone2007nonsaturated,daneshgaran2008unsaturated,dai2012unified,dai2014backoff}, in which all nodes communicate with a common receiver and lie within each other's carrier-sensing range. Under this fully connected topology, all nodes observe essentially the same channel state, which greatly simplifies the analysis. As a result, the behavior of single-AP IEEE 802.11 networks is relatively well understood. In particular, Bianchi's seminal Markov-chain model established the canonical saturation-throughput analysis of IEEE 802.11 DCF under ideal channel conditions~\cite{bianchi2000performance}, and subsequent studies extended this framework to non-saturated operation~\cite{malone2007nonsaturated}, non-ideal channels with capture effects~\cite{daneshgaran2008unsaturated}, and performance analysis and optimization in buffered WLANs~\cite{dai2012unified,dai2014backoff}. While these studies substantially improved the understanding of CSMA/CA in homogeneous single-cell WLANs, they mainly assess performance under given access parameters. To further reveal performance limits and optimal parameter settings, analytical models based on the access behavior of head-of-line (HOL) packets were developed~\cite{dai2012unified,dai2014backoff}.

When the network topology becomes more general, however, different links may observe different subsets of ongoing transmissions, and the access behavior of links becomes strongly coupled. This additional spatial heterogeneity makes throughput analysis substantially more difficult than in the single-AP case and has motivated extensive research on analytical modeling for wireless CSMA networks with general topologies. Existing approaches can be broadly classified into two categories: \emph{set-centric} and \emph{node-centric}.

In set-centric approaches, the global network behavior is characterized by sets of simultaneously active links. A seminal contribution in this direction is due to Boorstyn \emph{et al.}~\cite{boorstyn1987throughput}, who modeled a CSMA wireless network as a continuous-time Markov chain (CTMC) whose states correspond to independent sets of concurrently active links. Under saturated traffic and exponentially distributed backoff timers, they showed that the steady-state distribution admits a product-form solution. Building on this framework, Wang and Kar~\cite{wang2005throughput} and Durvy \emph{et al.}~\cite{durvy2009fairness} studied fairness issues in saturated CSMA networks, while Bellalta \emph{et al.}~\cite{bellalta2015interactions} and Faridi \emph{et al.}~\cite{faridi2016analysis} analyzed overlapping WLANs under static and dynamic channel bonding, respectively. More recently, Tarzjani and Krishnamachari~\cite{tarzjani2025computing} developed a Markov-chain-based computational approach for exact saturation-throughput characterization in heterogeneous \(p\)-CSMA networks over arbitrary conflict graphs. The back-of-the-envelope (BoE) method in~\cite{5198774}, derived from the ideal CSMA network model, further provides a simple throughput estimation rule.

Because set-centric approaches model the network at a global level, they can capture coupling across the topology. However, they typically rely on idealized assumptions, most notably the zero-propagation-delay assumption, under which simultaneous transmission attempts do not lead to collisions. This limitation becomes increasingly pronounced in dense networks, where non-negligible propagation delay and synchronous access attempts can substantially affect throughput. Moreover, many set-centric formulations are built on conflict-graph abstractions or aligned sensing/interference assumptions, and therefore do not directly capture the mismatch between carrier sensing and actual interference. Finally, these methods often suffer from high computational complexity because they require explicit enumeration or manipulation of large sets of feasible concurrent transmissions.

In node-centric approaches, each node or link is taken as the basic modeling unit, and its throughput is expressed as a function of the activities or throughputs of neighboring links. Ng and Liew~\cite{ng2007throughput} analyzed a single multihop flow over a linear chain, but their framework ignores collisions caused by simultaneous access attempts within the carrier-sensing range and is not readily extendable to arbitrary nonlinear topologies. Gao \emph{et al.}~\cite{gao2006determining} studied throughput characteristics under general network topologies, but relied on an exponential approximation for packet service time, which makes it difficult to accurately capture the non-memoryless service process induced by realistic backoff and collision dynamics. Jindal and Psounis~\cite{jindal2009achievable} characterized the achievable rate region of 802.11 multihop networks through a decompose-and-combine approach, which depends on the assumption that a complex topology can be decomposed into multiple bilateral sub-topologies and may introduce significant error in dense networks. For multi-BSS IEEE 802.11 networks with universal frequency reuse, Gao \emph{et al.}~\cite{7932180} grouped nodes according to spatial regions and established a discrete-time Markov renewal process to model HOL-packet behavior. However, their model assumes independence among channel-idle probabilities across APs and therefore neglects the coupling introduced by transmissions of nodes in overlapping regions. Garetto \emph{et al.}~\cite{garetto2005modeling} decomposed a multihop wireless network into embedded two-flow subgraphs, but multi-flow dependencies beyond pairwise interactions were not explicitly characterized.

Compared with set-centric methods, node-centric approaches avoid independent-set enumeration and therefore usually have lower complexity. However, because they do not explicitly model the network as a whole, they often rely on idealized assumptions, such as perfect carrier sensing, independence among transmitters, or simplified service processes, in order to keep the analysis tractable. These assumptions become increasingly inaccurate in dense multi-AP deployments, where local sensing differences and cross-AP coupling play a central role. As will be shown in Section~\ref{sect:case-study}, their accuracy degrades noticeably when link behaviors are highly correlated.

Another closely related line of research concerns the mismatch between carrier sensing and actual interference. Classical studies have long recognized that hidden-terminal and exposed-terminal effects can severely degrade CSMA/CA performance~\cite{kleinrock1975packet,jang2012hidden}. More recently, this issue has become increasingly important in dense WLAN deployments and spatial-reuse-oriented standards such as IEEE 802.11ax. For example, Wilhelmi \emph{et al.}~\cite{wilhelmi2021spatial} provided a comprehensive treatment of spatial reuse in IEEE 802.11ax, and Lanante and Roy~\cite{lanante2022obsspd} analyzed OBSS\_PD-based spatial reuse. These studies highlight the importance of sensing--interference mismatch, but they are largely tailored to specific mechanisms or deployment structures. A general analytical framework that directly captures arbitrary sensing and interference topologies remains largely unavailable.

The analytical model developed in this paper differs from the above lines of research in several key aspects. 
Compared with existing single-AP models~\cite{bianchi2000performance,malone2007nonsaturated,daneshgaran2008unsaturated,dai2012unified,dai2014backoff}, it is not restricted to fully connected contention scenarios with a common receiver and is applicable to wireless CSMA networks with arbitrary sensing and interference topologies. Compared with set-centric approaches~\cite{boorstyn1987throughput,wang2005throughput,durvy2009fairness,garetto2008modeling,nardelli2012closed}, it does not rely on the zero-propagation-delay assumption and therefore remains effective in dense-network regimes where simultaneous transmissions and collisions are non-negligible. Compared with node-centric approaches~\cite{ng2007throughput,gao2006determining,medepalli2006towards,jindal2009achievable,7932180}, it explicitly captures global coupling among links rather than relying on local independence or decoupling assumptions. Compared with studies tailored to specific manifestations of sensing--interference mismatch, such as hidden terminals or OBSS\_PD-based spatial reuse~\cite{jang2012hidden,wilhelmi2021spatial,lanante2022obsspd}, it provides a unified framework that directly models arbitrary sensing and interference topologies. As a result, the proposed framework offers a general analytical tool for throughput characterization and access-parameter optimization in strongly coupled wireless CSMA networks.

\section{System Model}
\label{sect:sys}

Consider a wireless network consisting of \(K\) links sharing the same spectrum, where each link is formed by a transmitter--receiver pair. Each transmitter employs carrier sense multiple access (CSMA) for medium access. Specifically, before transmitting a packet, transmitter \(i\) senses the channel and accesses the medium with probability \(q_i\) if the channel is sensed idle; otherwise, it defers transmission. Throughout this paper, we focus on the saturated regime, in which every transmitter always has packets available for transmission.

We consider a slotted CSMA network, where time is divided into equal-length slots. Each transmitter requires one slot to sense the channel and is allowed to initiate a transmission only at the beginning of a slot. In practical wireless CSMA networks, not all transmitters lie within each other's carrier-sensing range. The sensing relationship among links is therefore described by an undirected graph \(G_S(\mathcal{V},\mathcal{E}_S)\), referred to as the \emph{sensing graph}, where \(\mathcal{V}\) is the set of links and \((i,j)\in\mathcal{E}_S\) if the transmitters of links \(i\) and \(j\) can sense each other. For each link \(i\), let \(\mathcal{H}_i\) denote the set of links that can sense link \(i\), i.e.,
\begin{equation}
\mathcal{H}_i=\left\{j:(i,j)\in\mathcal{E}_S\right\}.
\end{equation}

At the receiver side, we adopt a noise-free channel and the classical collision model. Under this model, a packet transmitted on link \(i\) is successfully received if and only if no interfering link transmits concurrently. Let \(\mathcal{R}_i\) denote the set of links whose transmissions can interfere with link \(i\). For example, for the topology shown in Fig.~\ref{fig:network_topology}, we have \(\mathcal{R}_2=\{1,3\}\). We further assume that both successful and failed packet transmissions occupy \(\tau\) time slots\footnote{In CSMA systems, acknowledgement~(ACK) frames are used to indicate whether a packet transmission succeeds. Under the basic access mechanism, a successful transmission lasts slightly longer than a failed one because of the ACK exchange and the short inter-frame space. Since the ACK duration is typically much smaller than the payload-transmission duration, this difference is often negligible. In contrast, under the request-to-send/clear-to-send~(RTS/CTS) mechanism, the difference between successful and failed transmission durations can be significantly larger~\cite{bianchi2000performance}. Extending the model to unequal success and failure durations is left for future work.}.

The interference relationship among links is represented by a directed graph \(G_I(\mathcal{V},\mathcal{E}_I)\), referred to as the \emph{interference graph}, where \(\mathcal{E}_I\subseteq \mathcal{V}\times\mathcal{V}\). A directed edge \((i,j)\in\mathcal{E}_I\) indicates that the transmission of link \(i\) can interfere with the reception of link \(j\). Unlike the sensing graph, the interference graph is generally asymmetric because interference is determined by the transmitter--receiver geometry of the links.

The above model naturally captures both the hidden-terminal problem and the flow-in-the-middle problem. These two effects coexist in the topology shown in Fig.~\ref{fig:network_topology}. In particular, link~3 is outside the carrier-sensing range of link~2, yet its transmission can still interfere with the receiver of link~2, which gives rise to a hidden-terminal effect. At the same time, link~1 can sense both links~2 and~3, whereas links~2 and~3 cannot sense each other. Consequently, links~2 and~3 may repeatedly transmit in an overlapping manner, which prevents link~1 from gaining channel access and leads to the flow-in-the-middle phenomenon. Such topologies have a pronounced impact on throughput and fairness, and they will serve as representative scenarios in the subsequent analysis.

\section{Multi-Channel Modeling}
\label{sect:multichannel}

\subsection{Equivalent Multi-Channel Network}

In this subsection, we establish an equivalent multi-channel representation for a single-channel CSMA network. The basic intuition is as follows. If two links can sense each other, then they contend for access in a manner analogous to sharing the same channel. Conversely, if two links cannot sense each other, they behave as if they were operating on different logical channels. This observation motivates a graph-based transformation from the original CSMA network to an equivalent multi-channel network.

From the viewpoint of the sensing graph \(G_S\), each set of mutually sensing links forms a clique\footnote{In graph theory, a clique is a subset of vertices such that every pair of distinct vertices is connected by an edge.}. In this paper, we use maximal cliques\footnote{A maximal clique is a clique that cannot be extended by adding any adjacent vertex. The Bron--Kerbosch algorithm~\cite{10.1145/362342.362367} is a widely used method for enumerating all maximal cliques, and its complexity will be discussed in Section~\ref{sect:complex}.} to construct logical channels. Specifically, each maximal clique in \(G_S\) is mapped to one logical channel in the equivalent network. For the topology shown in Fig.~\ref{fig:network_topology}, links~1 and~2 form one maximal clique and are therefore regarded as sharing the first logical channel, whereas links~1 and~3 form another maximal clique and are regarded as sharing the second logical channel. Since link~1 belongs to both maximal cliques, its transmission occupies both logical channels simultaneously.

We now formalize this transformation. Consider a CSMA network whose sensing graph contains \(N\) maximal cliques. The network is equivalently represented as an \(N\)-channel network. For each link \(i\), let \(\mathcal{U}_i\) denote the set of maximal cliques to which it belongs. In the equivalent multi-channel network, link \(i\) is treated as a multi-channel device that adopts the following access rule: it can initiate transmission only when all channels in \(\mathcal{U}_i\) are idle, and once transmission starts, it occupies all channels in \(\mathcal{U}_i\) for \(\tau\) time slots.

It is important to emphasize that this equivalent multi-channel representation is determined solely by the sensing graph \(G_S\). The transformation provides the basis for the analytical framework developed next, in which the activity of the resulting multi-channel network is modeled explicitly.

\subsection{Multi-Channel Model}
\label{sect:model}

\begin{figure}[t]
  \centering
  \includegraphics[width=0.27\textwidth]{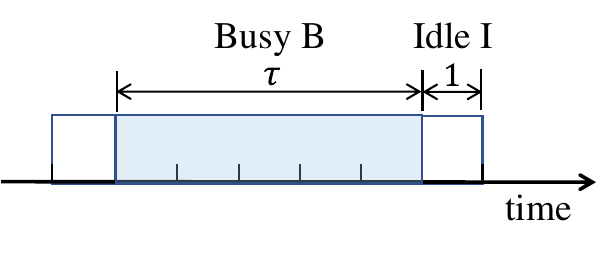}
  \caption{States of the logical channels.}
  \label{fig:channel_state}
\end{figure}

The channel dynamics of the equivalent multi-channel network are modeled by a discrete-time Markov renewal process \(\left(\bm{\mathcal{X}},\boldsymbol{V}\right)=\left\{\left(\boldsymbol{X}_j,V_j\right),j=0,1,\ldots\right\}\),
where \(V_j\) denotes the epoch of the \(j\)-th state transition and \(\boldsymbol{X}_j\) denotes the network state immediately after that transition. The state \(\boldsymbol{X}_j\) is defined as
\begin{equation}
\boldsymbol{X}_j=\Big(
\underbrace{X_j^{(1)},X_j^{(2)},\dots,X_j^{(N)}}_{\boldsymbol{X}_j^s},
\underbrace{D_j^{(1)},\dots,D_j^{(N-1)}}_{\boldsymbol{X}_j^c}
\Big),
\label{eq:state_xj}
\end{equation}
which consists of two components, namely \(\boldsymbol{X}_j^s\) and \(\boldsymbol{X}_j^c\). The first component, \(\boldsymbol{X}_j^s\), describes the states of the \(N\) logical channels, where \(X_j^{(i)}\) denotes the state of channel \(i\) after the \(j\)-th transition. As shown in Fig.~\ref{fig:channel_state}, each channel has two possible states: busy (\(\mathrm{B}\)) and idle (\(\mathrm{I}\)). The busy state lasts for \(\tau_{\mathrm{B}}=\tau\) time slots, whereas the idle state lasts for \(\tau_{\mathrm{I}}=1\) time slot.

\begin{figure}[t]
  \centering
  \includegraphics[width=0.4\textwidth]{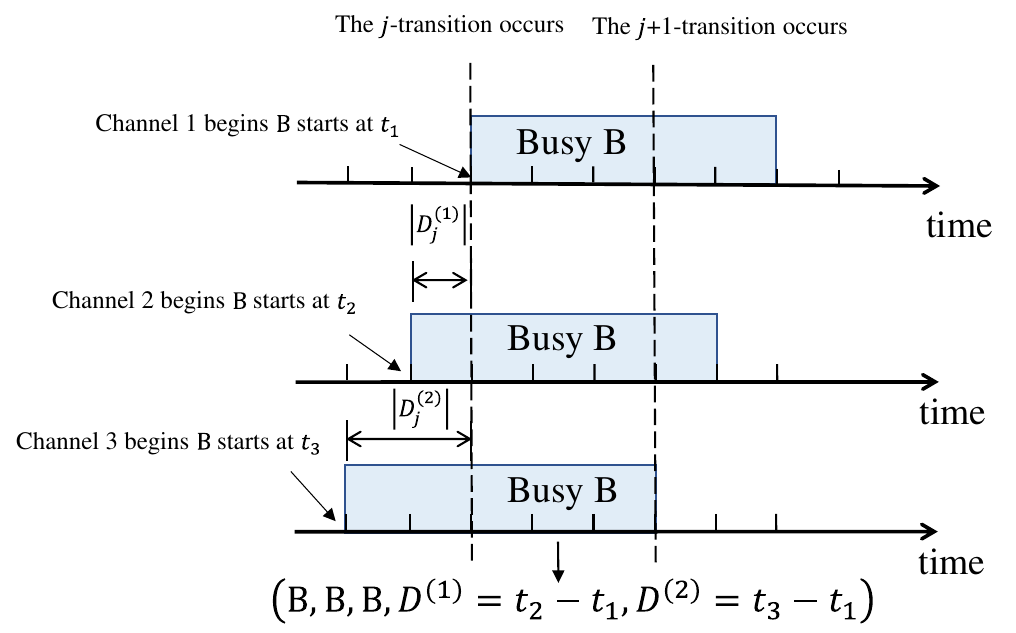}
  \caption{Definition and calculation of \(D_j^{(i)}\).}
  \label{fig:Calculation_A}
\end{figure}

When a link belonging to multiple maximal cliques starts transmitting, all corresponding logical channels become busy simultaneously. Therefore, the state transitions of different channels are generally coupled rather than independent. To capture this coupling, the second component \(\boldsymbol{X}_j^c\) is introduced as a set of time-offset variables. Specifically, \(D_j^{(i)}\) is defined as the difference between the starting times of the states of channel \(i+1\) and channel~1 at the \(j\)-th transition. Fig.~\ref{fig:Calculation_A} illustrates the definition and calculation of \(D_j^{(i)}\).

Let \(\mathbb{S}\) denote the state space of \(\bm{\mathcal{X}}\), and let \(p_{\boldsymbol{\nu},\boldsymbol{\mu}}\) denote the one-step transition probability from state \(\boldsymbol{\nu}\in\mathbb{S}\) to state \(\boldsymbol{\mu}\in\mathbb{S}\). Algorithm~1 outlines the procedure for constructing these transition probabilities.

\begin{algorithm}[t!]
    \caption{Determination of state-transition probabilities}
    \KwIn{The set of states of interest \(\mathcal{I}\)}
    \KwOut{The transition probabilities \(p_{\boldsymbol{\mu},\boldsymbol{\nu}}\) of the embedded Markov chain}
    Initialize \(p_{\boldsymbol{\mu},\boldsymbol{\nu}} \gets 0\) for all \(\boldsymbol{\mu},\boldsymbol{\nu}\in\mathbb{S}\)\;
    \For{each state \(\boldsymbol{\mu}\in\mathcal{I}\)}{
        Determine the set of links eligible for transmission in state \(\boldsymbol{\mu}\), denoted by \(\mathcal{K}^{\boldsymbol{\mu}}\)\;
        Enumerate all feasible link-transmission combinations from \(\mathcal{K}^{\boldsymbol{\mu}}\)\;
        \For{each feasible transmission combination}{
            Compute its occurrence probability \(p\)\;
            Determine the next state \(\boldsymbol{\nu}\)\;
            \(p_{\boldsymbol{\mu},\boldsymbol{\nu}} \gets p_{\boldsymbol{\mu},\boldsymbol{\nu}} + p\)\;
        }
    }
    \label{alg:transition_prob}
\end{algorithm}

The steady-state distribution of the embedded Markov chain \(\bm{\mathcal{X}}=\{\boldsymbol{X}_j\}\) is obtained from
\begin{equation}
\left\{
\begin{array}{l}
\pi_{\boldsymbol{\nu}}=\sum_{\boldsymbol{\mu}\in\mathbb{S}} p_{\boldsymbol{\mu},\boldsymbol{\nu}}\pi_{\boldsymbol{\mu}},\\
1=\sum_{\boldsymbol{\nu}\in\mathbb{S}} \pi_{\boldsymbol{\nu}},
\end{array}
\right.
\label{eq:steady_embedded}
\end{equation}
where \(\pi_{\boldsymbol{\nu}}\) denotes the steady-state probability of state \(\boldsymbol{\nu}\) in the embedded chain.

The limiting state probabilities of the Markov renewal process \(\left(\bm{\mathcal{X}},\boldsymbol{V}\right)\) are then given by
\begin{equation}
\tilde{\pi}_{\boldsymbol{\mu}}
=
\frac{\pi_{\boldsymbol{\mu}}\tau_{\boldsymbol{\mu}}}
{\sum\limits_{\boldsymbol{\nu}\in\mathbb{S}}\pi_{\boldsymbol{\nu}}\tau_{\boldsymbol{\nu}}},
\label{eq:limiting_probs}
\end{equation}
where \(\tau_{\boldsymbol{\mu}}\) is the holding time of state \(\boldsymbol{\mu}\in\mathbb{S}\), measured in time slots. For a state
\[
\left(
X^{(1)},X^{(2)},\dots,X^{(N)},D^{(1)},\dots,D^{(N-1)}
\right),
\]
the holding time is
\begin{equation}
\begin{mysmall}
\begin{aligned}
&\tau_{\left( X^{(1)}, X^{(2)}, \dots, X^{(N)}, D^{(1)}, \dots, D^{(N-1)} \right)} \\
&= \min\big( \tau_{X^{(1)}}, \tau_{X^{(2)}} + D^{(1)}, \dots, \tau_{X^{(N)}} + D^{(N-1)} \big) \\
&\quad - \max\big( 0, D^{(1)}, \dots, D^{(N-1)} \big).
\end{aligned}
\end{mysmall}
\end{equation}

Given the sensing graph \(G_S\), the limiting state probabilities can be expressed explicitly as functions of the transmission probabilities \(q\) and the packet-transmission duration \(\tau\) by using~(\ref{eq:steady_embedded}) and~(\ref{eq:limiting_probs}). As an illustrative example, consider the sensing graph in Fig.~\ref{fig:network_topology}. The limiting state probabilities are
\begin{equation}
\begin{aligned}
\tilde{\pi}_{(\mathrm{I},\mathrm{I},0)}
&=
\frac{1}
{D}, \\
\tilde{\pi}_{(\mathrm{I},\mathrm{B},k)}
&=
\frac{(1-q_1)q_2}{D}, \quad k=0,\ldots,1-\tau,
\\
\tilde{\pi}_{(\mathrm{B},\mathrm{I},k)}
&=
\frac{(1-q_1)q_3}{D}, \quad k=0,\ldots,\tau-1.
\end{aligned}
\end{equation}
respectively, where $D = 1 + (1 - q_1)q_2 q_3 \tau^2
+ \big(1 + (q_2 + q_3 - 1)(1 - q_1)\big)\tau$. The detailed derivation is provided in Appendix~\ref{app:limiting-probs}.

Based on these explicit expressions of the limiting state probabilities, the throughput of each link can be derived in a closed form in the next section.

\section{Link Throughput}
\label{sect:thp}

In this section, we derive the throughput of each link as an explicit function of the transmission probabilities \(q\).

\subsection{Link Throughput}

The throughput of link \(i\) is given by
\begin{equation}
\label{eq:thp_main}
\hat{\lambda}_i
=
\tau \sum_{\boldsymbol{\mu}\in \mathcal{Y}_i}
q_i \, \phi_i(\boldsymbol{\mu}) \, \tilde{\pi}_{\boldsymbol{\mu}},
\end{equation}
where \(\mathcal{Y}_i\) denotes the set of network states in which link \(i\) is eligible to transmit, namely,
\begin{equation}
\mathcal{Y}_i
=
\left\{
\boldsymbol{X}\in\mathbb{S}:
X^{(j)}=\mathrm{I}, \ \forall j\in\mathcal{U}_i
\right\}.
\end{equation}
Here, \(\phi_i(\boldsymbol{\mu})\), referred to as the \emph{interruption-free probability}, is the probability that a transmission initiated by link \(i\) at time \(t\) completes successfully without interruption, given that the network state at time \(t-1\) is \(\boldsymbol{\mu}\). The term \(\tilde{\pi}_{\boldsymbol{\mu}}\) is the limiting probability that the network is in state \(\boldsymbol{\mu}\) at time \(t-1\).

The interruption-free probability can be computed by enumerating all interference-free transmission evolutions and evaluating their probabilities. To illustrate this computation, consider the topology in Fig.~\ref{fig:network_topology} and focus on link~2. Assume that \(\tau=3\). Since link~2 belongs only to the first logical channel, it is eligible to transmit whenever that channel is idle. Therefore,
\begin{equation}
\begin{aligned}
\mathcal{Y}_{2}
=
\{(\mathrm{I},\mathrm{I},0),\,
(\mathrm{I},\mathrm{B},0),(\mathrm{I},\mathrm{B},-1),\,
(\mathrm{I},\mathrm{B},-2)\}.
\end{aligned}
\end{equation}
According to the interference graph in Fig.~\ref{fig:network_topology}, links~1 and~3 can interrupt the transmission of link~2. Hence, a successful transmission of link~2 requires that neither link~1 nor link~3 transmits during the entire transmission interval. We next evaluate \(\phi_2(\boldsymbol{\mu})\) for each state in \(\mathcal{Y}_2\).

\begin{enumerate}
	\item \emph{State \((\mathrm{I},\mathrm{I},0)\):} When the network is in state \((\mathrm{I},\mathrm{I},0)\), all links are eligible to transmit in the next slot. The probability that neither link~1 nor link~3 transmits is \((1-q_1)(1-q_3)\). If only link~2 starts transmitting, the network moves to state \((\mathrm{B},\mathrm{I},0)\).

	In state \((\mathrm{B},\mathrm{I},0)\), only link~3 still senses an idle channel and may transmit in the next slot. The probability that link~3 remains silent is \((1-q_3)\), which leads to state \((\mathrm{B},\mathrm{I},-1)\). Repeating the same argument over the remaining slots of the transmission yields
	\begin{equation}
	\label{eq:phi_case1}
	\phi_2(\mathrm{I},\mathrm{I},0)
	=
	(1-q_1)(1-q_3)^3.
	\end{equation}

	\item \emph{States \((\mathrm{I},\mathrm{B},0)\) and \((\mathrm{I},\mathrm{B},-1)\):} In either of these states, link~3 is already transmitting and will remain active in the next slot. Therefore, any transmission initiated by link~2 is immediately subject to interference from link~3, and hence
	\begin{equation}
	\label{eq:phi_case2}
	\phi_2(\mathrm{I},\mathrm{B},D)=0,
	\qquad
	D\in\{0,-1\}.
	\end{equation}

	\item \emph{State \((\mathrm{I},\mathrm{B},-2)\):} In this state, the ongoing transmission of link~3 ends in the next slot. The network then enters state \((\mathrm{B},\mathrm{I},0)\). To ensure that link~2 completes its transmission without interruption, link~3 must remain silent during the following two slots. Therefore,
	\begin{equation}
	\label{eq:phi_case3}
	\phi_2(\mathrm{I},\mathrm{B},-2)
	=
	(1-q_3)^2.
	\end{equation}
\end{enumerate}

\begin{figure}[t]
  \centering
  \includegraphics[width=0.4\textwidth]{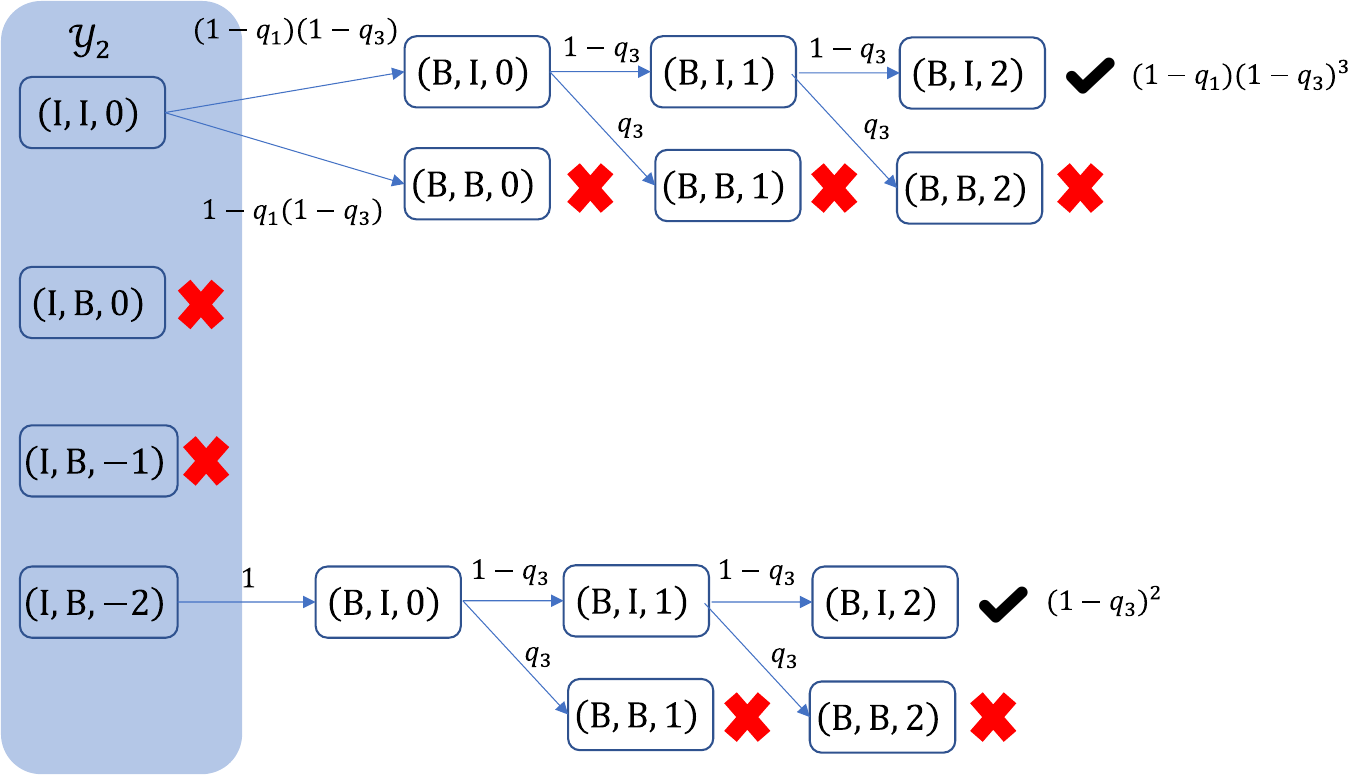}
  \caption{Illustration of the calculation of the probability that a transmission experiences no interruption during its entire duration.}
  \label{fig:Calculation_p}
\end{figure}

The above calculation is illustrated graphically in Fig.~\ref{fig:Calculation_p}. Substituting~(\ref{eq:phi_case1})--(\ref{eq:phi_case3}) into~(\ref{eq:thp_main}), the throughput of link~2 is obtained as
\begin{equation}
\begin{aligned}
\label{eq:throughput_link2}
\hat{\lambda}_2
=
\tau q_2 \Big(
&(1-q_1)(1-q_3)^3 \tilde{\pi}_{(\mathrm{I},\mathrm{I},0)}
\\
&+(1-q_3)^2 \tilde{\pi}_{(\mathrm{I},\mathrm{B},-2)}
\Big).
\end{aligned}
\end{equation}
Since the limiting state probabilities \(\tilde{\pi}\) have already been expressed as explicit functions of \(q\) in Section~\ref{sect:model}, the throughput of link~2 can be further written explicitly as
\begin{equation}
\label{eq:throughput_link2_q}
\begin{aligned}
\hat{\lambda}_2
=
\frac{3 q_2 (1 - q_1)(1 - q_3)^2 (1 + q_2 - q_3)}
{1 + 9(1 - q_1)q_2 q_3
 + 3\big(1 + (q_2 + q_3 - 1)(1 - q_1)\big)}.
\end{aligned}
\end{equation}

The same procedure applies to other links and other sensing/interference topologies. More importantly, the above derivation has a recursive structure and can therefore be implemented algorithmically. In particular, the interruption-free probability can be computed via a depth-first search, as summarized in Algorithm~2.

\begin{algorithm}[t!]
    \caption{Depth-first search for exact throughput evaluation}
    \KwIn{Target link \(i\)}
    \KwOut{Throughput \(\hat{\lambda}_i\)}
    Initialize \(\hat{\lambda}_i \gets 0\)\;

    \SetKwFunction{DFS}{DFS}
    \SetKwProg{Fn}{Function}{}{}

    \Fn{\DFS{\(\boldsymbol{\mu}, i\)}}{
        Initialize \(P_{\mathrm{no}}^{\boldsymbol{\mu},i} \gets 0\)\;
        \uIf{the transmission of link \(i\) has ended in state \(\boldsymbol{\mu}\)}{
            \Return \(1\)\;
        }
        \Else{
            Find \(\mathcal{K}^{\boldsymbol{\mu}}\), the eligible links in state \(\boldsymbol{\mu}\)\;
            Enumerate all feasible transmission combinations from \(\mathcal{K}^{\boldsymbol{\mu}}\)\;
            Select the subset \(\mathcal{S}_{i}^{\boldsymbol{\mu}}\) in which no interrupting link transmits\;
            \For{each transmission combination in \(\mathcal{S}_{i}^{\boldsymbol{\mu}}\)}{
                Compute its occurrence probability \(p\)\;
                Determine the next state \(\boldsymbol{\nu}\)\;
                \(P_{\mathrm{no}}^{\boldsymbol{\mu},i}
                \gets
                P_{\mathrm{no}}^{\boldsymbol{\mu},i}
                + p \cdot \DFS(\boldsymbol{\nu}, i)\)\;
            }
            \Return \(P_{\mathrm{no}}^{\boldsymbol{\mu},i}\)\;
        }
    }

    \For{each state \(\boldsymbol{\mu}\in \mathcal{Y}_i\)}{
        Compute \(P_{\mathrm{nt}}^{\boldsymbol{\mu},i}\), the probability that no link in \(\mathcal{R}_i\) is transmitting\;
        \If{\(P_{\mathrm{nt}}^{\boldsymbol{\mu},i}\neq 0\)}{
            Find \(\mathcal{K}^{\boldsymbol{\mu}}\), the eligible links in state \(\boldsymbol{\mu}\)\;
            Enumerate all feasible transmission combinations from \(\mathcal{K}^{\boldsymbol{\mu}}\)\;
            Select the subset \(\mathcal{S}_{i}^{\boldsymbol{\mu}}\) in which link \(i\) transmits and no interrupting link transmits\;
            \For{each transmission combination in \(\mathcal{S}_{i}^{\boldsymbol{\mu}}\)}{
                Compute its occurrence probability \(p\)\;
                Determine the next state \(\boldsymbol{\nu}\)\;
                \(\hat{\lambda}_i
                \gets
                \hat{\lambda}_i
                + P_{\mathrm{nt}}^{\boldsymbol{\mu},i}\, p\, \DFS(\boldsymbol{\nu}, i)\, \tilde{\pi}_{\boldsymbol{\mu}}\)\;
            }
        }
    }
    \label{alg:dfs_exact}
\end{algorithm}

\subsection{Link Throughput Approximation}
\label{sect:link_thp_approx}

The exact evaluation of~(\ref{eq:thp_main}) requires computing the interruption-free probability over the entire transmission interval. Although Algorithm~\ref{alg:dfs_exact} yields an exact characterization by explicitly tracking the network evolution after transmission begins, the resulting computational complexity can be high. As pointed out in~\cite{tinnirello2009rethinking}, there is generally a tradeoff between modeling accuracy and complexity.

To reduce the computational burden\footnote{The resulting complexity reduction is discussed in Section~\ref{sect:complex}.}, we derive a lower bound on the interruption-free probability and use it as a throughput approximation:
\begin{equation}
\begin{aligned}
\phi_i(\boldsymbol{\mu})
&\ge
\prod_{j\in\mathcal{R}_i}
\Pr\Big\{
\text{link } j \text{ does not transmit during } \tau-g_j^{\boldsymbol{\mu}}
\Big\} \\
&=
\prod_{j\in\mathcal{R}_i}
(1-q_j)^{\tau-g_j^{\boldsymbol{\mu}}},
\end{aligned}
\end{equation}
where \(g_j^{\boldsymbol{\mu}}\) is the minimum number of time slots required for all channels sensed by link \(j\) to become idle when the network is in state \(\boldsymbol{\mu}\).

Again consider link~2 in the topology of Fig.~\ref{fig:network_topology}, with \(\tau=3\). According to the interference graph, both links~1 and~3 can interrupt the transmission of link~2. To illustrate the lower bound, consider the contribution of link~3. Since link~3 belongs only to the second maximal clique, it senses only the second logical channel. When the network is in state \((\mathrm{I},\mathrm{I},0)\), the second channel is already idle, and hence \(g_3^{(\mathrm{I},\mathrm{I},0)}=0\). Therefore, for link~2 to succeed, link~3 must remain silent for the next \(\tau-g_3^{(\mathrm{I},\mathrm{I},0)}=3\) slots, which gives the factor \((1-q_3)^{\tau-g_3^{(\mathrm{I},\mathrm{I},0)}}\).

When the network is in state \((\mathrm{I},\mathrm{B},-2)\), the second channel becomes idle in the next slot, and thus \(g_3^{(\mathrm{I},\mathrm{B},-2)}=1\). In this case, the probability that link~3 remains silent for the subsequent \(\tau-g_3^{(\mathrm{I},\mathrm{B},-2)}=2\) slots is at least \((1-q_3)^{\tau-g_3^{(\mathrm{I},\mathrm{B},-2)}}\).

Accordingly, a lower bound on the throughput of link~2 is
\begin{equation}
\begin{mysmall}
\begin{aligned}
\hat{\lambda}_2
\ge
\hat{\lambda}_2^{\mathrm{LB}}
=
\tau q_2 \Big(
(1-q_1)^3(1-q_3)^3 \tilde{\pi}_{(\mathrm{I},\mathrm{I},0)}
+
(1-q_3)^2 \tilde{\pi}_{(\mathrm{I},\mathrm{B},-2)}
\Big).
\end{aligned}
\end{mysmall}
\end{equation}

The corresponding approximation procedure is summarized in Algorithm~3.

\begin{algorithm}[htb]
    \caption{Approximate evaluation of link throughput}
    \KwIn{Target link \(i\)}
    \KwOut{Lower bound \(\hat{\lambda}_i^{\mathrm{LB}}\)}
    Initialize \(\hat{\lambda}_i^{\mathrm{LB}}\gets 0\)\;
    \For{each state \(\boldsymbol{\mu}\in\mathcal{Y}_i\)}{
        Compute \(P_{\mathrm{nt}}^{\boldsymbol{\mu},i}\), the probability that no link in \(\mathcal{R}_i\) is transmitting\;
        \If{\(P_{\mathrm{nt}}^{\boldsymbol{\mu},i}\neq 0\)}{
            Compute \(g_j^{\boldsymbol{\mu}}\) for each \(j\in\mathcal{R}_i\)\;
            \(
            \hat{\lambda}_i^{\mathrm{LB}}
            \gets
            \hat{\lambda}_i^{\mathrm{LB}}
            +
            P_{\mathrm{nt}}^{\boldsymbol{\mu},i}
            \prod_{j\in\mathcal{R}_i}(1-q_j)^{\tau-g_j^{\boldsymbol{\mu}}}
            \tilde{\pi}_{\boldsymbol{\mu}}
            \)\;
        }
    }
    \label{alg:approx_thp}
\end{algorithm}

\subsection{Special Case: Interference Graph as a Subgraph of the Sensing Graph}

We next consider a special case in which the interference graph is a subgraph of the sensing graph. A typical practical example is the request-to-send/clear-to-send (RTS/CTS) mechanism, which is widely used to mitigate the hidden-terminal problem. Under RTS/CTS, before transmitting a data packet, the transmitter first sends an RTS frame, and the intended receiver responds with a CTS frame. Nodes that hear either the RTS or the CTS defer their own transmissions for the announced duration. As a result, hidden terminals are prevented from interrupting an ongoing transmission once the RTS/CTS exchange has been successfully completed.

Under this mechanism, the effective interference relationship is restricted by the sensing relationship, and the interference graph \(G_I\) becomes a subgraph of the sensing graph \(G_S\). Indeed, if link \(i\) can interfere with link \(j\), then the transmitter of link \(i\) lies within the reception range of the receiver of link \(j\). Hence, when link \(j\) successfully initiates transmission, the transmitter of link \(i\) can hear the CTS sent by the receiver of link \(j\) and will defer its own access.

Note, however, that RTS/CTS does not eliminate collisions entirely. Due to non-negligible propagation delay, two links within hearing range may still initiate transmission simultaneously, causing a collision at the beginning of the transmission. The key effect of RTS/CTS is that collisions can only occur at transmission initiation, rather than during an ongoing transmission. Therefore, the interruption-free probability in~(\ref{eq:thp_main}) is greatly simplified and can be written as
\begin{equation}
\label{eq:phi_rtscts}
\phi_i(\boldsymbol{\mu})
=
\prod_{j\in \mathcal{K}^{\boldsymbol{\mu}}\cap \mathcal{R}_i}
(1-q_j),
\end{equation}
where \(\mathcal{K}^{\boldsymbol{\mu}}\cap \mathcal{R}_i\) denotes the set of links that are both eligible to transmit in state \(\boldsymbol{\mu}\) and capable of interfering with link \(i\) at transmission initiation.

Again consider link~2 in the topology of Fig.~\ref{fig:network_topology}. Under RTS/CTS, the hidden-terminal interference from link~3 to link~2 is removed, and only collisions occurring at transmission initiation need to be considered. Since Appendix~\ref{app:limiting-probs} expresses the limiting state probabilities \(\tilde{\pi}\) explicitly as functions of the packet duration \(\tau\), the throughput of link~2 can be written as an explicit function of both \(\tau\) and \(q\):
\begin{equation}
\begin{aligned}
\hat{\lambda}_{2}
&=
\tau q_2 \big(
(1-q_1)\tilde{\pi}_{(\mathrm{I},\mathrm{I},0)}
+\tilde{\pi}_{(\mathrm{I},\mathrm{B},0)}
+\tilde{\pi}_{(\mathrm{I},\mathrm{B},-1)}
+\tilde{\pi}_{(\mathrm{I},\mathrm{B},-2)}
\big)
\\
&=
\frac{\tau q_2 (1 - q_1)(1 + \tau q_2)}
{1 + \tau^{2} (1 - q_1) q_2 q_3
 + \tau \big(1 + (q_2 + q_3 - 1)(1 - q_1)\big)}.
\end{aligned}
\end{equation}

\section{Case Study}
\label{sect:case-study}

In this section, we apply the proposed analytical framework to two representative CSMA scenarios: multi-BSS IEEE 802.11 networks with universal frequency reuse, and ad-hoc networks. These two scenarios cover both infrastructure-based and decentralized wireless networks and highlight the capability of the proposed model to handle strong spatial coupling, hidden terminals, exposed terminals, and flow-in-the-middle effects.


\subsection{Multi-BSS IEEE 802.11 Networks with Universal Frequency Reuse}

\begin{figure}[t]
  \centering
  \includegraphics[width=0.52\textwidth,height=13cm,keepaspectratio]{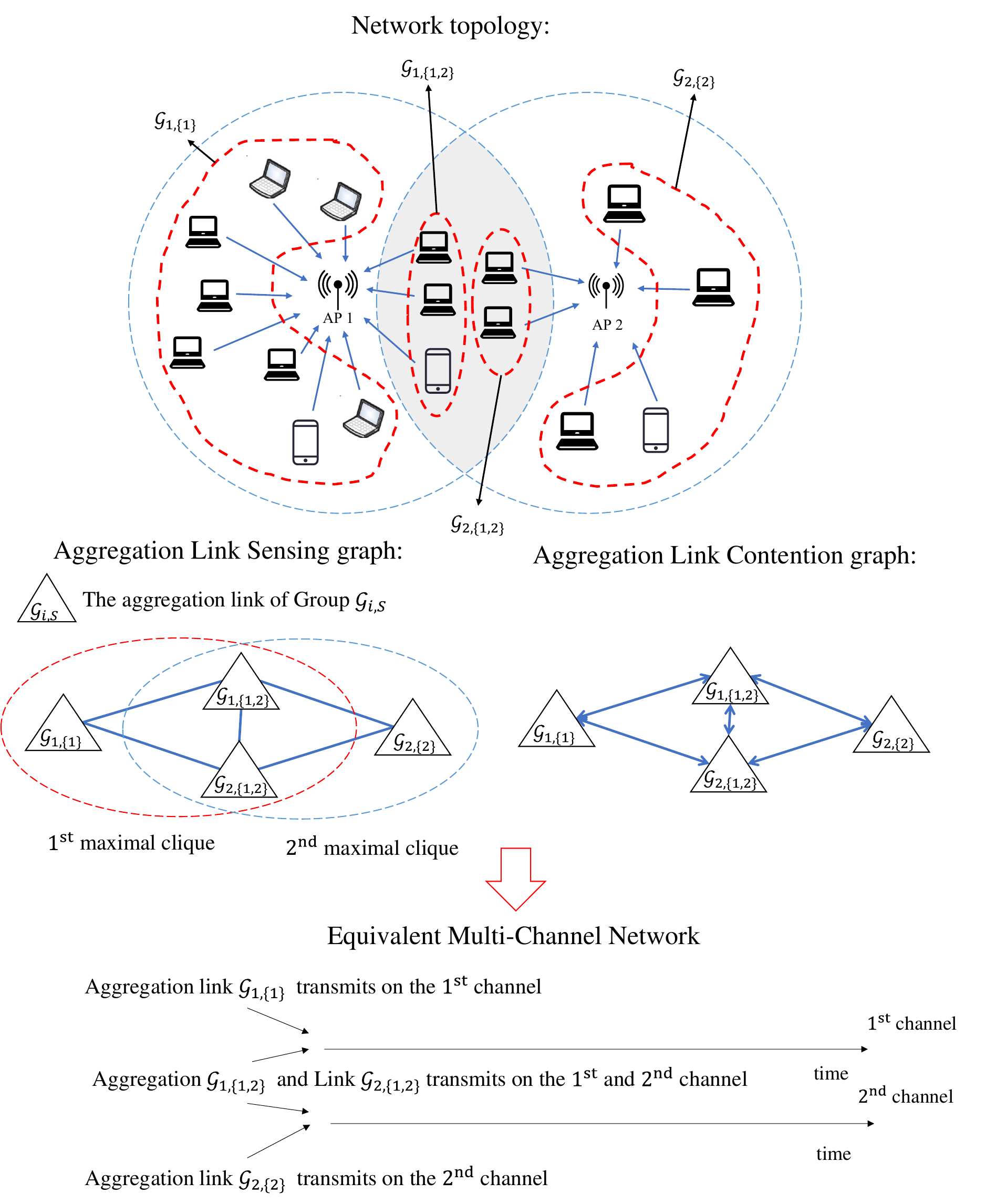}
  \caption{Illustration of link aggregation.}
   \label{fig:multi-bss_trans}
\end{figure}

\subsubsection{System Model}

An important application of CSMA networks is IEEE 802.11, in which the network consists of multiple basic service sets (BSSs). Each BSS has a star topology centered at an access point (AP), and multiple stations access the medium through the distributed coordination function (DCF). Due to the limited availability of unlicensed spectrum, multiple BSSs often operate on the same frequency band, i.e., under universal frequency reuse. In such deployments, nodes at different locations may observe different carrier-sensing environments. In particular, as illustrated in Fig.~\ref{fig:multi-bss_trans}, nodes located in overlapping coverage regions of multiple BSSs can sense transmissions from more than one BSS and therefore experience a lower probability of finding the channel idle. This gives rise to strong coupling among the medium-access behaviors of different BSSs.

Consider the uplink of an IEEE 802.11 DCF network with \(M\) BSSs, where multiple stations in each BSS transmit to a common AP. Each station therefore forms one link with its associated AP, and all BSSs share the same spectrum. We define the coverage of a BSS as the region centered at its AP within which the AP can hear the transmissions of all associated nodes. Since the coverage regions of different BSSs may overlap, some nodes may be heard by multiple APs and associate with one of them.

For carrier sensing, we assume that nodes infer channel availability from AP feedback rather than relying solely on direct sensing. In practice, this feedback can be realized through RTS/CTS: after receiving an RTS frame, the AP broadcasts a CTS frame indicating the channel occupation duration, and all nodes that hear the AP defer their transmissions accordingly, even if they cannot directly hear the transmitting node. Under this mechanism, nodes in overlapping coverage regions can obtain channel availability information from multiple APs, while their own transmissions may affect not only the associated BSS but also neighboring BSSs.

To characterize channel occupancy, let \(\tau_T\) denote the duration of a successful transmission, measured in slots. For IEEE 802.11, \(\tau_T\) can be written as~\cite{zhang2022synchronous}
\begin{equation}
\begin{aligned}
\tau_T
=
\Bigg(
&\frac{L_P+\text{MAC header}}{R}
+\text{SIFS}
+\frac{\text{ACK}}{\text{Basic Rate}}
\\
&+\text{DIFS}
+\text{PHY preamble}
\Bigg)\Big/\sigma,
\end{aligned}
\label{tauTcal}
\end{equation}
where \(\sigma\) is the slot duration. We further assume that the channel occupancy in collision, denoted by \(\tau_F\), satisfies \(\tau_F=\tau_T=\tau\). This corresponds to a conservative setting in which a collision occupies the channel for the same duration as a successful transmission.

Let \(\mathcal{M}=\{1,2,\dots,M\}\) denote the set of BSSs. Since a node may be heard by multiple APs, we group the links according to two attributes: the BSS \(i\) with which the node is associated, and the subset \(\mathcal{S}\subseteq\mathcal{M}\) of APs that can hear that node. The corresponding group is denoted by \(G_{i,\mathcal{S}}\). Let \(|G_{i,\mathcal{S}}|=n^{(i,\mathcal{S})}\) be the number of nodes in this group, and let \(q_i\) denote the transmission probability of a node associated with BSS \(i\).

Links in the same group \(G_{i,\mathcal{S}}\) have identical sensing and interference relationships and can sense one another. Therefore, all links in the same group can be aggregated into a single equivalent link. As illustrated in Fig.~\ref{fig:multi-bss_trans}, after aggregating nodes into group-level links, the multi-BSS IEEE 802.11 network can be represented as an equivalent multi-channel network. The throughput analysis then follows directly from the framework developed in Section~\ref{sect:multichannel}. Let \(\hat{\lambda}^{A}_{(i,\mathcal{S})}\) denote the throughput of the aggregation link corresponding to group \(G_{i,\mathcal{S}}\). The throughput of an individual node in that group is given by
\begin{equation}
\hat{\lambda}_{(i,\mathcal{S})}
=
\frac{q_i(1-q_i)^{n^{(i,\mathcal{S})}-1}}
{1-(1-q_i)^{n^{(i,\mathcal{S})}}}
\hat{\lambda}^{A}_{(i,\mathcal{S})},
\end{equation}
where the prefactor is the probability that, conditioned on a successful transmission of the aggregation link, exactly one node in the group transmits. The total network throughput is then given by
\begin{equation}
\hat{\lambda}_{\mathrm{out}}
=
\sum_{i\in\mathcal{M}}
\sum_{\mathcal{S}\subseteq\mathcal{M}}
n^{(i,\mathcal{S})}\hat{\lambda}_{(i,\mathcal{S})}.
\end{equation}


\subsubsection{Simulation Results}

We next validate the proposed analysis through event-driven simulation. In all simulations, each transmitter is saturated. A transmitter selects its backoff counter uniformly from \(\{0,1,\dots,W_i\}\), where \(W_i\) is the backoff window, senses the channel at the beginning of each slot, and decrements its counter by one whenever the channel is sensed idle. A transmission starts when the backoff counter reaches zero. The simulated throughput is computed as \(\tau N/T\), where \(N\) is the number of successfully transmitted packets over an observation interval of \(T=10^9\) slots. The system parameter setting follows \cite{zhang2022synchronous}, except that the payload length is set to \(L_P = 18432\) bits, with \(\tau\) computed according to \eqref{tauTcal}, yielding \(\tau = 27\).

We first consider the two-BSS topology shown in Fig.~\ref{fig:multi-bss_trans}. In this network, nodes are divided into four groups:
\(G_{1,\{1\}}\),
\(G_{2,\{2\}}\),
\(G_{1,\{1,2\}}\), and
\(G_{2,\{1,2\}}\).
Nodes in \(G_{i,\{i\}}\) belong to BSS \(i\) and can be heard only by AP \(i\), whereas nodes in \(G_{i,\{1,2\}}\) belong to BSS \(i\) but are within the coverage of both AP~1 and AP~2. For each node in BSS \(i\), the backoff window is \(W_i\), and the corresponding transmission probability after sensing an idle channel is \(q_i=2/(W_i+1)\)~\cite{dai2012unified}.

To benchmark the proposed model, we compare it with three representative baselines.

\begin{itemize}
\item \emph{HOL:} We use the analytical model in~\cite{gao2017throughput} as the representative node-centric baseline. In that model, a discrete-time Markov renewal process is established for the head-of-line (HOL) packet of a node in group \(G_{i,\mathcal{S}}\). The service rate of a node in that group is
\begin{equation}
\begin{aligned}
\tilde{\pi}_T^{(i,\mathcal{S})}
=
\frac{\alpha^{(i,\mathcal{S})}\tau}
{\alpha^{(i,\mathcal{S})}\left(\tau+\tau\frac{1-p^{(i)}}{p^{(i)}}\right)+\frac{1}{p^{(i)}}\cdot\frac{1+W_i}{2}},
\end{aligned}
\label{eq:service_rate}
\end{equation}
where \(\alpha^{(i,\mathcal{S})}\) is the probability that a node in group \(G_{i,\mathcal{S}}\) senses the channel idle, and \(p^{(i)}\) is the probability of successful transmission of a HOL packet in BSS \(i\).

\item \emph{CTMC:} We use the model in~\cite{nardelli2012closed} as the representative set-centric baseline. It characterizes the network by the independent sets of simultaneously active links and is known to be accurate when collisions caused by simultaneous transmission attempts are negligible.

\item \emph{BOE:} We also consider the Back-of-the-Envelope (BOE) method in~\cite{5198774}. BOE computes normalized link throughput from the maximum independent sets (MISs) of the contention graph. Specifically, if \(n^{\mathrm{BOE}}\) is the number of MISs and \(n_i^{\mathrm{BOE}}\) is the number of MISs containing link \(i\), then the normalized throughput of link \(i\) is \(n_i^{\mathrm{BOE}}/n^{\mathrm{BOE}}\).
\end{itemize}

\begin{figure*}[t]

    \centering
        \subfloat[]{
		\includegraphics[height=0.35\textwidth]{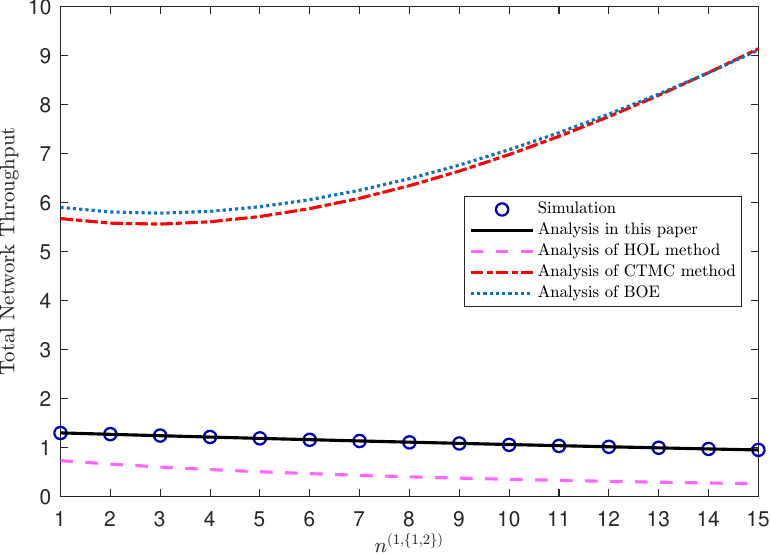}
		\label{fig:topo_BSS_n}
        }
	\centering
    	\subfloat[]{
		\includegraphics[height=0.35\textwidth]{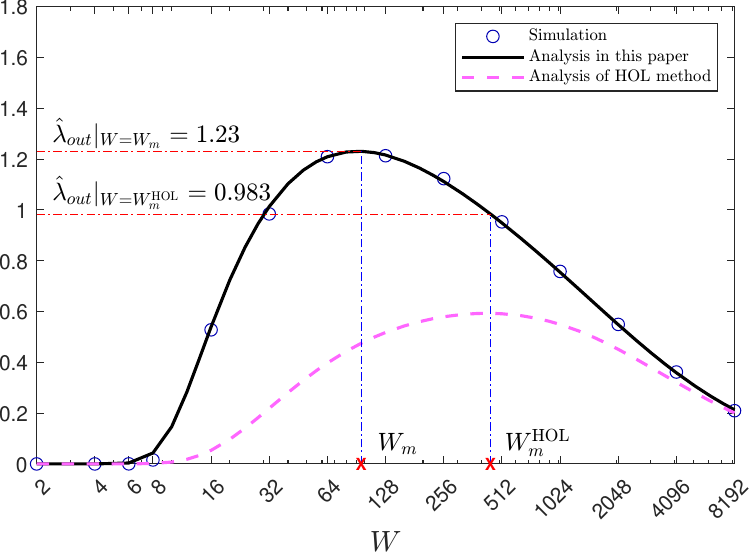}
		\label{fig:topo_BSS_1}
        }
    \caption{Total network throughput performance in a two-BSS network with \(\tau=27\). (a) Network throughput versus \(n^{(1,\{1,2\})}\) . \(W^{(1)}=W^{(2)}=32\), \(n^{(1,\{1\})}=n^{(2,\{2\})}=n^{(2,\{1,2\})}=4\) (b) Network throughput versus the backoff window \(W\).  \(W^{(1)}=W^{(2)}=W\), \(n^{(1,\{1\})}=n^{(2,\{2\})}=n^{(1,\{1,2\})}=n^{(2,\{1,2\})}=20\).}
	\label{fig:BSS_total_thp}
\end{figure*}

Fig.~\ref{fig:topo_BSS_n} shows how the network throughput changes with the number of nodes in the overlapping region, \(n^{(1,\{1,2\})}\). The proposed model consistently agrees well with simulation across the entire range of \(n^{(1,\{1,2\})}\). In contrast, the HOL baseline underestimates the throughput, while both the CTMC and BOE baselines significantly overestimate it.

The reasons for the loss of accuracy in the HOL and CTMC baselines are different. For the HOL model~\cite{gao2017throughput}, a key parameter is \(\alpha^{(i,\mathcal{S})}\), the probability that a node in \(G_{i,\mathcal{S}}\) senses the channel idle. To obtain a tractable model, this quantity is approximated as
\begin{equation}
\alpha^{(i,\mathcal{S})}
=
\prod_{j\in\mathcal{S}}
\alpha^{(j,\{j\})}.
\label{eq:alpha_gao}
\end{equation}
Equation~(\ref{eq:alpha_gao}) assumes that the idle-sensing events seen by different BSSs are independent. This approximation becomes increasingly inaccurate when transmissions in the overlapping region become more frequent, because those transmissions induce common busy periods observed by multiple BSSs and therefore introduce strong correlation in channel sensing.

The CTMC model~\cite{nardelli2012closed} tends to be optimistic, as shown in Fig.~\ref{fig:topo_BSS_n}. Although it captures contention interactions at the network-state level, it still relies on a continuous-time abstraction with independent exponential backoff timers. This approximation becomes inaccurate under high contention, where synchronized transmission attempts are no longer negligible, and therefore leads to overly optimistic throughput predictions. The BOE method also deviates noticeably from simulation in Fig.~\ref{fig:topo_BSS_n}. This is expected because BOE is derived from the ideal CSMA network model, which assumes zero propagation delay and zero duration for failed transmissions, whereas both effects are non-negligible in the simulations considered here. By contrast, the proposed model maintains high accuracy in dense overlapping regions and under high transmission probability. This is because the model represents the network state jointly across multiple logical channels and explicitly includes time-domain variables that capture their coupling. As a result, the correlation induced by shared overlapping nodes is preserved rather than approximated away.

Fig.~\ref{fig:topo_BSS_1} shows the total network throughput as a function of the backoff window \(W\) in a dense overlapping region with \(n^{(1,\{1,2\})}=n^{(2,\{1,2\})}=20\). The figure indicates that the backoff window has a strong impact on network performance: both overly small and overly large values of \(W\) reduce the total throughput, and hence careful tuning is necessary. Indeed, the optimal backoff window predicted by the proposed model, denoted by \(W_m\), differs substantially from that obtained by the HOL model, denoted by \(W_m^{\mathrm{HOL}}\). The throughput achieved by using \(W_m\) yields a 25.1\% improvement over that obtained using \(W_m^{\mathrm{HOL}}\). These results show that the proposed model can accurately characterize multi-BSS performance across a wide range of node densities and can provide more reliable guidance for backoff-window optimization. \footnote{The CTMC baseline is omitted due to its prohibitive computational complexity in dense scenarios.}

\subsection{Ad-hoc Networks}

\begin{figure}[t]
	\centering
	\includegraphics[width=0.3\textwidth]{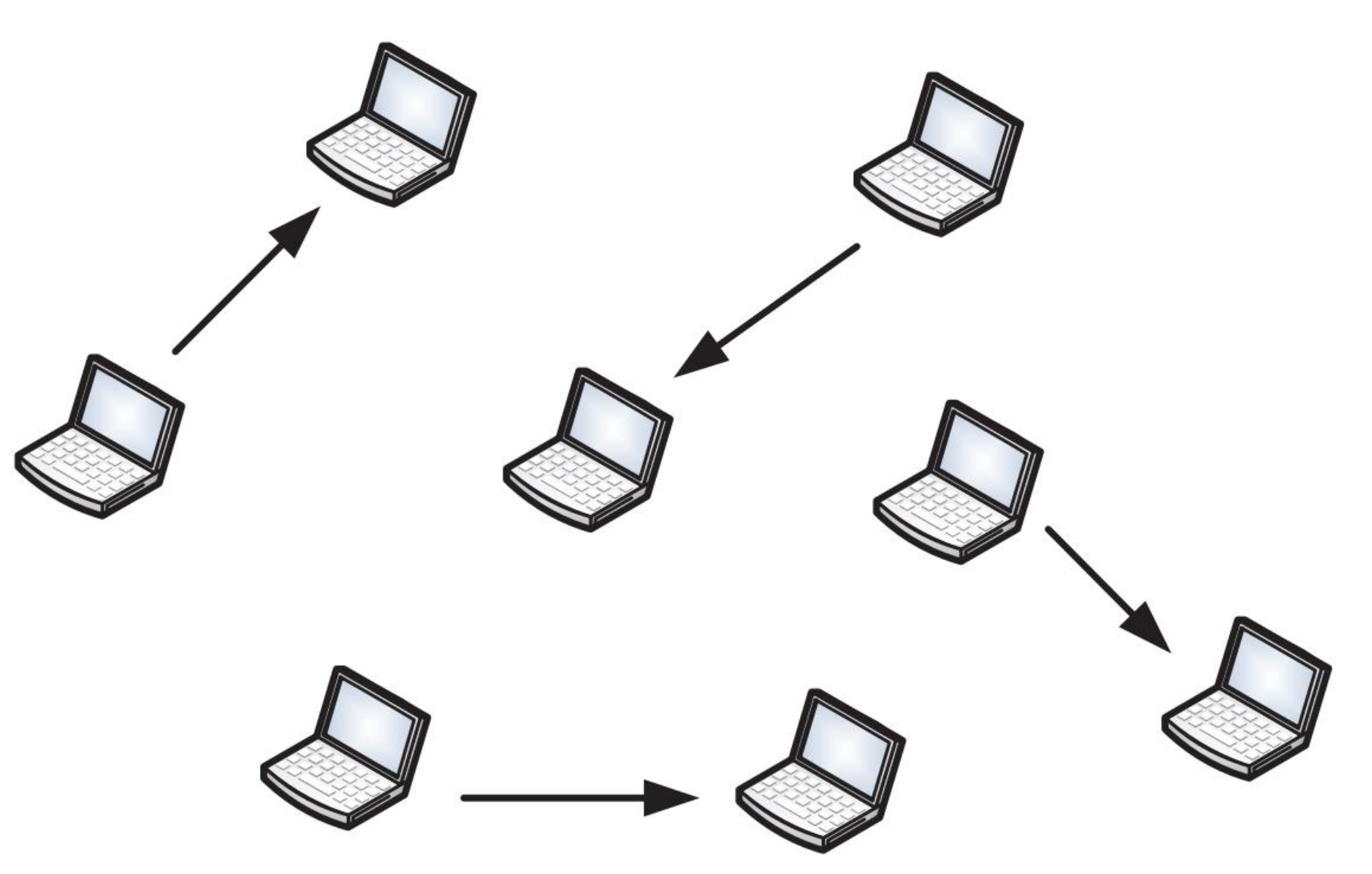}
	\caption{Illustration of ad-hoc scenarios.}
	\label{fig:ad-hoc}
\end{figure}

Another important class of CSMA networks is the ad-hoc network. As illustrated in Fig.~\ref{fig:ad-hoc}, an ad-hoc network consists of multiple autonomous nodes that communicate directly with one another without centralized infrastructure. Due to the absence of coordinating entities such as APs, ad-hoc networks are more vulnerable to interference-related phenomena. In this subsection, we consider a representative topology that simultaneously exhibits the hidden-terminal problem, the exposed-terminal problem, and the flow-in-the-middle effect. These phenomena are known to degrade network performance by increasing collisions and introducing unfair access opportunities. The system assumptions are the same as those in Section~\ref{sect:sys} and are not repeated here.

\begin{figure*}[t]
	\centering
	\includegraphics[width=0.8\textwidth]{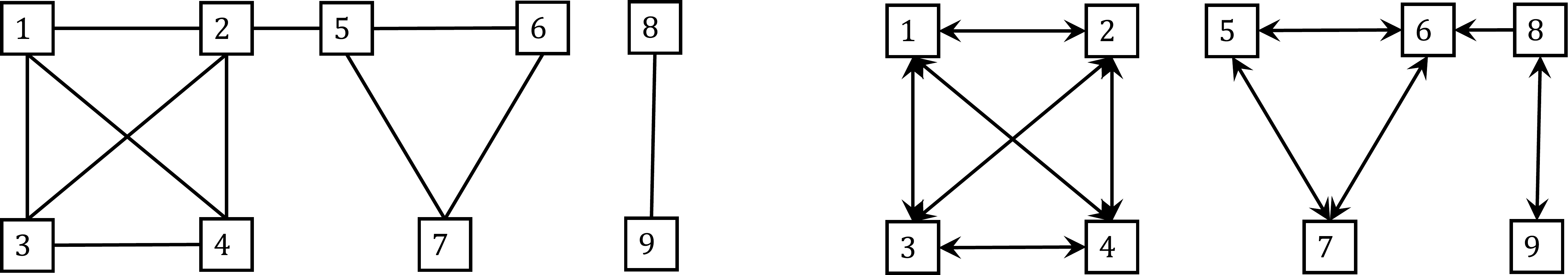}
	\caption{Sensing graph and interference graph of an ad-hoc network with nine links.}
	\label{fig:both_topo}
\end{figure*}

\begin{figure*}[t]
	\centering
    	\subfloat[]{
		\includegraphics[height=0.28\textwidth]{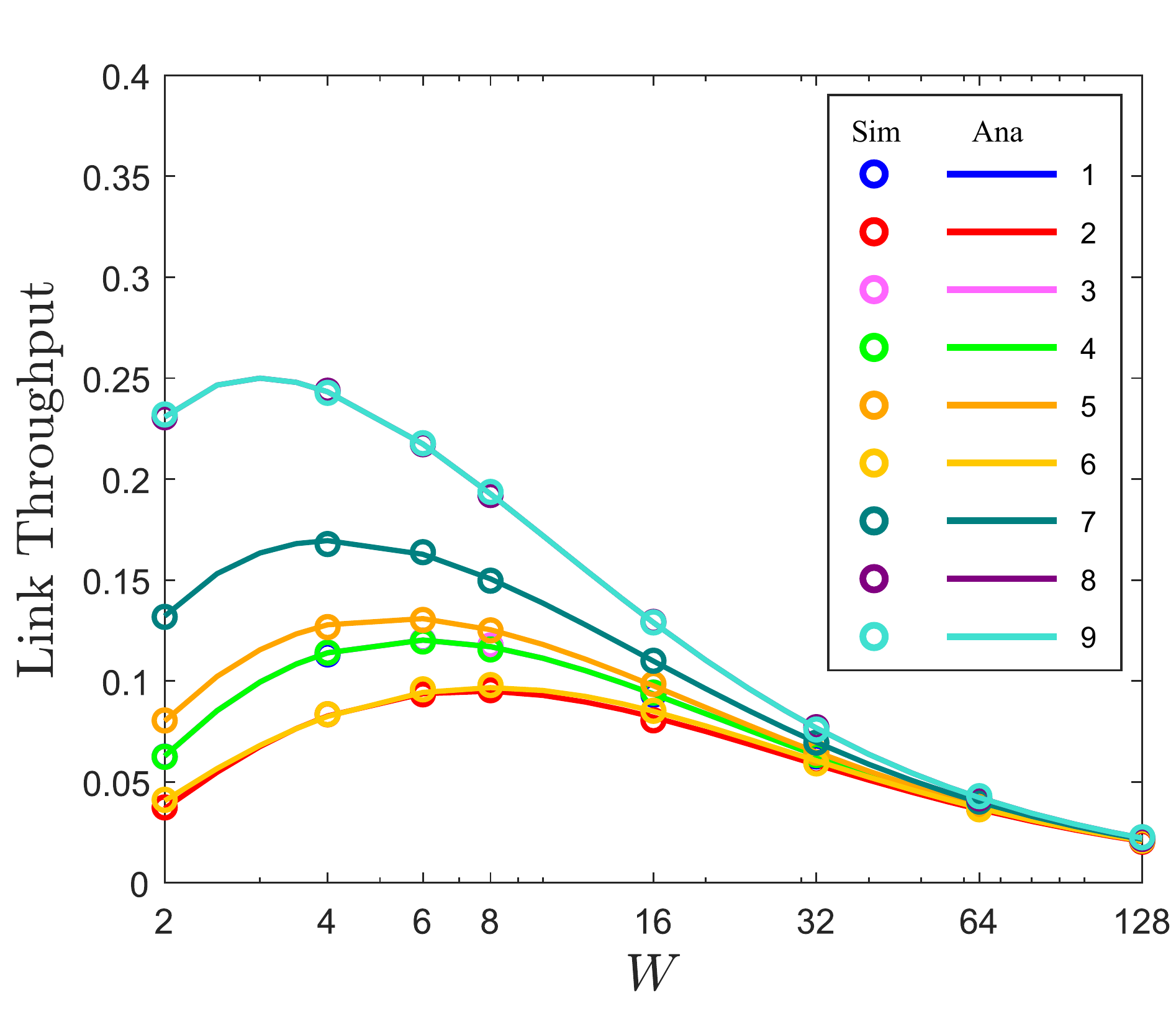}
		\label{fig:both_thp_1}}
	\subfloat[]{
		\includegraphics[height=0.28\textwidth]{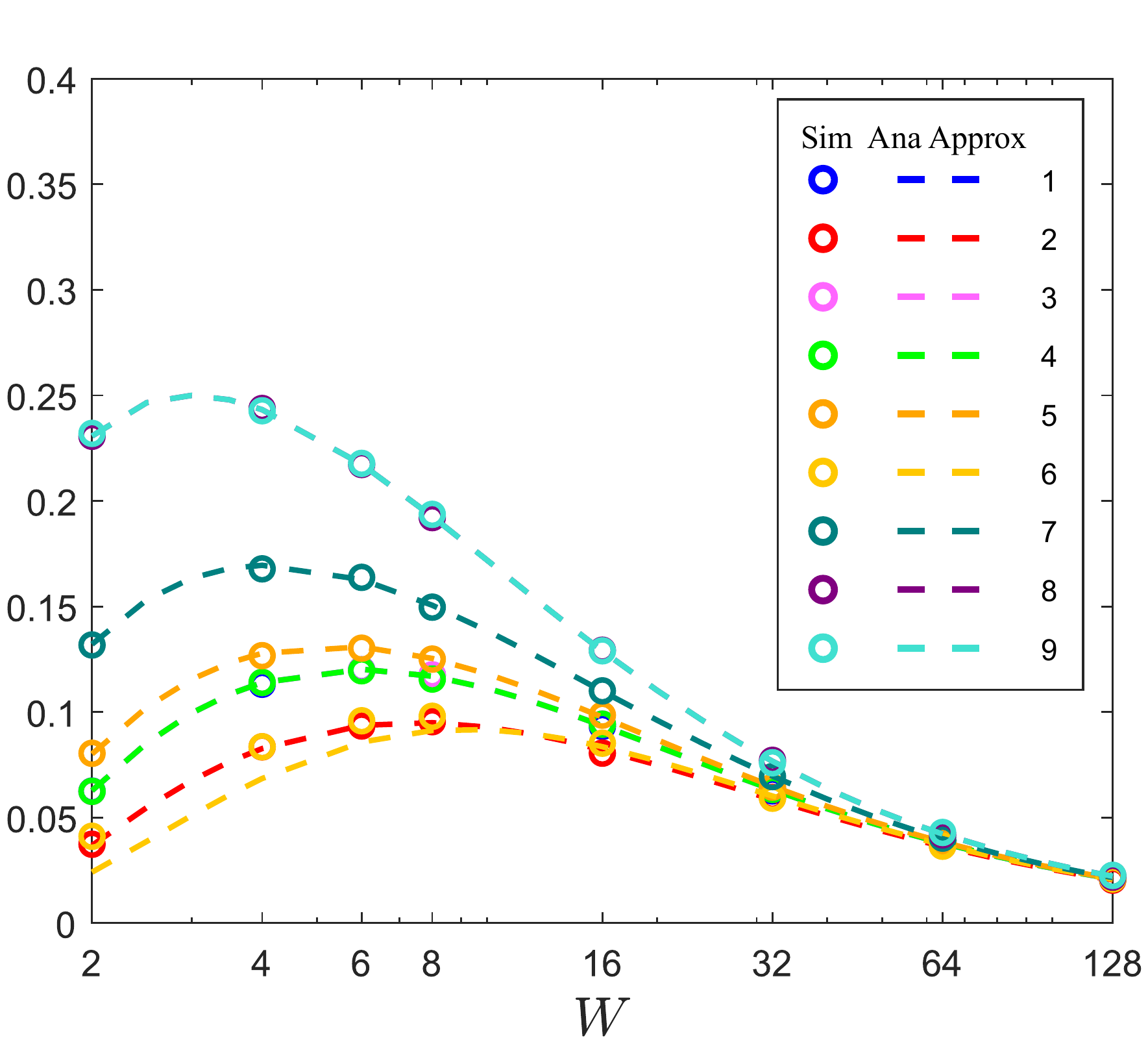}
		\label{fig:both_thp_2}}
    \subfloat[]{
		\includegraphics[height=0.28\textwidth]{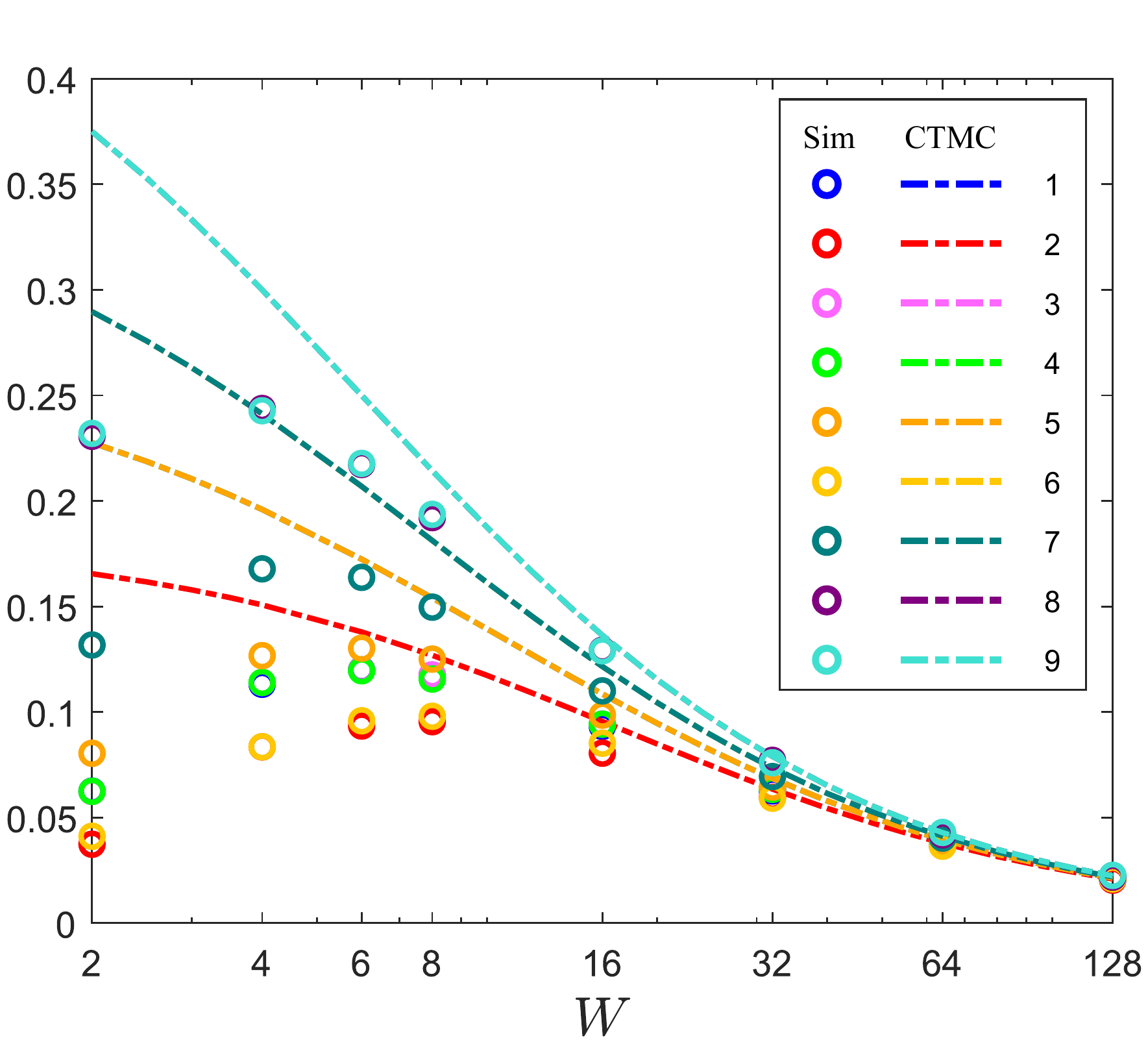}
		\label{fig:both_thp_3}}
    \caption{Throughput performance of the ad-hoc network in Fig.~\ref{fig:both_topo} with \(\tau=3\).}
	\label{fig:both_thp}
\end{figure*}

\begin{figure}[t]
	\centering
    \includegraphics[width=0.45\textwidth]{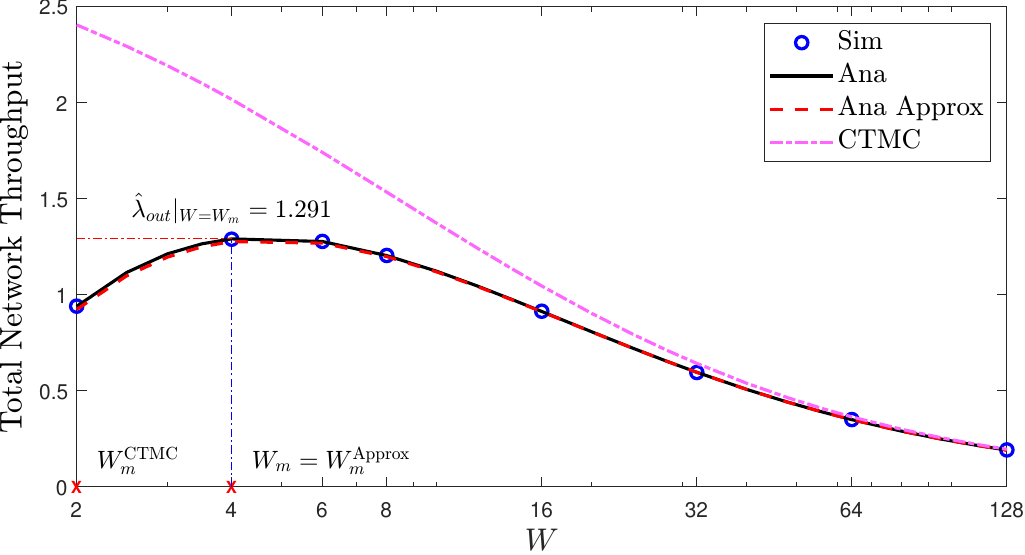}
	\caption{Total network throughput for the topology in Fig.~\ref{fig:both_topo}, with \(\tau=3\).}
	\label{fig:9topo}
\end{figure}

We evaluate the nine-link ad-hoc topology shown in Fig.~\ref{fig:both_topo}. This topology simultaneously exhibits hidden-terminal, exposed-terminal, and flow-in-the-middle effects. In particular, link~6 cannot sense link~8, while the transmission of link~8 can still interfere with link~6, which gives rise to a hidden-terminal effect. In addition, link~2 can sense link~5 even though link~5 does not interfere with the reception of link~2, corresponding to an exposed-terminal effect. Meanwhile, in the sensing graph, link~2 is coupled with links~1,~3,~4, and~5, and thus serves as the flow-in-the-middle link in this topology.

In Fig.~\ref{fig:both_thp_1}, the exact analysis is compared with simulations, whereas Fig.~\ref{fig:both_thp_2} presents the lower-bound approximation together with the simulation results. The exact analysis matches the simulated throughput of all links closely over the entire range of \(W\). The approximation is also accurate for most links, and its visible deviation is concentrated on link~6, i.e., the link affected by the hidden terminal. In this case, the approximate result is slightly lower than the exact one, but it still remains close to the simulation. In Fig.~\ref{fig:both_thp_3}, the CTMC baseline is compared with simulations. As \(W\) decreases and the transmission probability increases, the CTMC prediction departs increasingly from the simulated results because collisions caused by simultaneous transmission attempts are not properly captured. Moreover, the CTMC baseline fails to distinguish the asymmetric hidden-terminal impact on links~6 and~7, and therefore predicts the same throughput for these two links, whereas the simulation shows that link~6 achieves lower throughput than link~7 due to the interference from link~8. Overall, the proposed model captures both nonzero-delay collisions and hidden-terminal effects, and therefore provides a much more accurate characterization of per-link throughput in this strongly coupled ad-hoc topology.

Fig.~\ref{fig:9topo} further shows the total network throughput for the same topology. We compare the optimal backoff window obtained from the proposed exact model, denoted by \(W_m\), the CTMC baseline, denoted by \(W_m^{\mathrm{CTMC}}\), and the approximation, denoted by \(W_m^{\mathrm{Approx}}\). Because the predicted throughput of the CTMC model keeps increasing as \(W\) decreases, and the resulting optimization would incorrectly favor an excessively small backoff window. In other words, based on the CTMC analysis, the optimal \(W\) would be driven toward the smallest possible value, which is clearly inconsistent with the simulation results. By contrast, the proposed model captures the throughput degradation caused by collisions in the small-\(W\) regime and therefore provides a more accurate characterization of the system. More importantly, the optimal backoff window predicted by the approximation coincides with that predicted by the exact analysis, which indicates that the approximation is effective for parameter optimization even when hidden terminals are present.

\section{Computational Complexity Analysis}
\label{sect:complex}

In the previous sections, we showed that the proposed analytical framework can derive explicit throughput expressions for a given CSMA topology. In this section, we analyze its computational complexity. The overall complexity arises from three main components: the construction of the equivalent multi-channel network, the solution of the Markov steady-state equations, and the evaluation of the interruption-free probability in the presence of hidden terminals.

The first source of complexity comes from transforming the original CSMA network into the equivalent multi-channel network. This step requires identifying all maximal cliques in the sensing graph \(G_S\). To this end, we employ the Bron--Kerbosch algorithm, which recursively explores candidate cliques while pruning infeasible branches, as summarized in Algorithm~\ref{alg:bronkerbosch_basic}. The worst-case time complexity of this algorithm is \(O(3^{K/3})\)~\cite{10.1145/362342.362367}, where \(K\) is the number of vertices in the graph. On the other hand, for certain sparse graph families, such as planar graphs and graphs with low arboricity, the number of maximal cliques grows only linearly with the number of vertices. In such cases, all maximal cliques can be enumerated in linear time~\cite{chiba1985arboricity,chrobak1991planar}.

\begin{algorithm}[t]
	\caption{Bron--Kerbosch Algorithm}
	\KwIn{Graph \(G_S(\mathcal{V},\mathcal{E}_S)\)}
	\KwOut{All maximal cliques in \(G_S\)}
	Initialize \(R \gets \emptyset\), \(P \gets \mathcal{V}\), \(X \gets \emptyset\)\;
	\SetKwFunction{BronKerbosch}{BronKerbosch}
	\SetKwProg{Fn}{Function}{}{}
	
	\Fn{\BronKerbosch{\(R,P,X\)}}{
		\If{\(P=\emptyset\) \textbf{and} \(X=\emptyset\)}{
			Output \(R\) as a maximal clique\;
		}
		\ForEach{\(v\in P\)}{
			\BronKerbosch{\(R\cup\{v\},\,P\cap N(v),\,X\cap N(v)\)}\;
			\(P \gets P\setminus\{v\}\)\;
			\(X \gets X\cup\{v\}\)\;
		}
	}
	\BronKerbosch{\(R,P,X\)}\;
	\label{alg:bronkerbosch_basic}
\end{algorithm}

The second component of complexity arises from solving the steady-state linear system associated with the embedded Markov chain, namely~(\ref{eq:steady_embedded}). If this linear system is solved by Gaussian elimination, the computational complexity is \(O(n^3)\), where \(n\) is the number of variables, equivalently the number of network states. Before simplification, the number of states is on the order of $2^N \tau^{2(N-1)}$,
where \(N\) is the number of logical channels, i.e., the number of maximal cliques, and \(\tau\) is the packet-transmission duration. Therefore, this part of the complexity increases exponentially with both the number of channels and the packet duration. For example, when \(\tau=4\) and \(N=4\), the total number of states is 8192.

However, by applying the state-space reduction procedure described in Appendix~\ref{app:limiting-probs}, the number of variables can be reduced substantially. After simplification, the number of effective variables becomes approximately \(\sum_{i=1}^{N} \binom{N}{i}\), which corresponds to the number of states in \(\mathcal{I}\) and depends only on the number of channels. For example, when \(N=5\), the number of reduced states is only 31. This reduction greatly improves the tractability of the steady-state analysis.

The third source of complexity appears in scenarios where hidden terminals are present. In this case, one must compute the interruption-free probability appearing in~(\ref{eq:thp_main}). The complexity of this computation grows exponentially with \(\tau\), because all non-interfering transmission evolutions over the \(\tau\)-slot transmission interval must be enumerated and their probabilities accumulated. The resulting complexity is \(O\!\left(\tau\,2^K\,2^N\,\tau^{2(N-1)}\right)
\), where \(2^K\) accounts for all possible link-transmission combinations, and \(2^N\tau^{2(N-1)}\) accounts for all possible network states.

The key reason for this complexity is that the exact computation requires a recursive exploration of future network states throughout the entire transmission interval. For each state, all feasible link-transmission combinations must be examined in order to determine whether the tagged transmission remains uninterrupted. By contrast, the lower-bound approximation introduced in Section~\ref{sect:link_thp_approx} avoids this state-by-state recursion and instead estimates the interruption-free probability directly from the earliest channel-access opportunities of the interfering links. Consequently, neither the detailed dynamics over the last \(\tau\) slots nor the exhaustive set of link-transmission combinations needs to be enumerated. The resulting computational complexity is reduced to \(O(KN)\).

Overall, the proposed framework incurs higher complexity than simplified node-centric approximations, but this additional cost enables it to capture global coupling, nonzero-delay collisions, and hidden-terminal effects more accurately. Moreover, the state-space reduction and lower-bound approximation substantially improve the computational tractability of the framework, especially when the number of maximal cliques in the sensing graph and the packet transmission duration are not excessively large.

\section{Conclusion}
\label{sect:conclusion}
In this paper, we developed a new analytical framework for throughput characterization in wireless CSMA networks with arbitrary sensing and interference topologies. By explicitly modeling sensing and interference through two distinct graphs, and by transforming the original CSMA network into an equivalent multi-channel network modeled via a discrete-time Markov renewal process, the proposed framework captures global coupling among links and enables explicit throughput characterization.

The proposed analysis was applied to several representative CSMA scenarios, including multi-BSS IEEE 802.11 networks with universal frequency reuse, and ad-hoc topologies exhibiting hidden-terminal, exposed-terminal, and flow-in-the-middle effects. Simulation results showed that the proposed model achieves more accurate throughput estimation than existing analytical approaches, especially in dense deployments and in scenarios with strong coupling among link behaviors. The resulting explicit expressions also provide useful insight into access-parameter optimization, such as backoff-window selection. Finally, we analyzed the computational complexity of the framework and discussed practical complexity-reduction methods.

\appendices

\section{Explicit Functions of the Limiting State Probabilities}
\label{app:limiting-probs}

In Section~\ref{sect:model}, the limiting state probabilities were expressed as explicit functions of the transmission probabilities \(q\). In this appendix, we further derive these probabilities as explicit functions of the packet-transmission duration \(\tau\).

A main difficulty is that the number of equations in~(\ref{eq:steady_embedded}) depends on \(\tau\), which makes it difficult to solve the resulting linear system analytically and obtain closed-form expressions in terms of \(\tau\). To address this issue, we transform~(\ref{eq:steady_embedded}) into an equivalent system whose dimension is independent of \(\tau\).

We first rewrite~(\ref{eq:steady_embedded}) as a system involving only the limiting state probabilities of the states in \(\mathcal{I}\). Using the transformation procedure described in Appendix~\ref{app:transformation},~(\ref{eq:steady_embedded}) and~(\ref{eq:limiting_probs}) can be organized into the matrix form
\begin{equation}
\label{eq:idle_eq_set_general}
\mathbf{y}=\mathbf{A}\mathbf{y}+\mathbf{b},
\end{equation}
where \(\mathbf{y}\) is the vector of limiting state probabilities associated with the states in \(\mathcal{I}\), and \(\mathbf{b}\) is a constant vector. Specifically,
\begin{equation}
\mathbf{y}
=
\left[
\tilde{\pi}_{(\mathrm{I},\mathrm{I},\dots,\mathrm{I},0,\dots,0)},
\cdots
\right]^T
\label{eq:Y_matrix}
\end{equation}
and
\begin{equation}
\mathbf{b}
=
\left[
\frac{1}{\tau},0,\cdots,0
\right]^T.
\end{equation}

It can be observed that, for different values of \(\tau\), some states in \(\mathcal{I}\) have identical limiting state probabilities. This property can be formalized as follows.

For a state \(\mu\in\mathbb{S}\), let \(\mathbb{I}_\mu\) denote the set of indices of idle channels:
\begin{equation}
\mathbb{I}_\mu
=
\left\{
i:X^{(i)}=\mathrm{I},\ i\in\{1,\dots,N\}
\right\}.
\end{equation}
For the busy channels, group together the indices of channels that become busy simultaneously. Let these subsets be denoted by \(\mathcal{C}_1,\mathcal{C}_2,\dots\), and collect them into the set
\begin{equation}
\mathbb{B}_\mu
=
\left\{
\mathcal{C}_1,\mathcal{C}_2,\dots
\right\}.
\end{equation}
Then, for two states \(\mu,\nu\in\mathbb{S}\), if
\[
\mathbb{I}_\mu=\mathbb{I}_\nu
\qquad\text{and}\qquad
\mathbb{B}_\mu=\mathbb{B}_\nu,
\]
we have
\[
\tilde{\pi}_\mu=\tilde{\pi}_\nu.
\]

By merging states with identical limiting state probabilities, the system~(\ref{eq:idle_eq_set_general}) can be compressed into a reduced system whose dimension is independent of \(\tau\). Consequently, the limiting state probabilities can be derived explicitly as functions of both \(\tau\) and \(q\).

We next illustrate this reduction procedure using the topology in Fig.~\ref{fig:multi-bss_trans}. Following Appendix~\ref{app:transformation},~(\ref{eq:steady_embedded}) can first be transformed into a system involving only the limiting state probabilities of the states in \(\mathcal{I}\), namely
\begin{equation}
\label{eq:idle_eq_set_example}
\mathbf{y}=\mathbf{A}\mathbf{y}+\mathbf{b},
\end{equation}
where
\begin{equation}
\begin{mysmall}
\begin{aligned}
\mathbf{y}
=
\Big[
\tilde{\pi}_{(\mathrm{I},\mathrm{I},0)},
\tilde{\pi}_{(\mathrm{I},\mathrm{B},0)},
\cdots,
\tilde{\pi}_{(\mathrm{I},\mathrm{B},1-\tau)},
\tilde{\pi}_{(\mathrm{B},\mathrm{I},0)},
\cdots,
\tilde{\pi}_{(\mathrm{B},\mathrm{I},\tau-1)}
\Big]^T
\end{aligned}
\end{mysmall}
\end{equation}
and
\begin{equation}
\mathbf{b}
=
\left[
\frac{1}{\tau},0,\cdots,0
\right]^T.
\end{equation}
The coefficient matrix \(\mathbf{A}\) is shown in Fig.~\ref{fig:matrix_A}.

\begin{figure*}[!t]
\centering
\begin{equation}
\begin{mysmall}
 \begin{aligned}
\mathbf{A}=\begin{bNiceMatrix}
 \rho_1\rho_{2}\rho_{3}-\frac{1}{\tau}&g(1)&\cdots & g(\tau-1) & \rho_{2}+g(\tau) & h(1) &\cdots  & h(\tau-1)& \rho_{3}+h(\tau)\\
\rho_1\rho_{2}(1-\rho_{3}) & 0 & &0 & 0 & 0&  &  0& 1-\rho_{3} \\
0 & \rho_{2} & \cdots& 0& 0 & 0 &  \cdots & 1-\rho_{3}& 0\\
  & \vdots &\ddots & \vdots& \vdots & \vdots & \iddots & \vdots&\\
0 & 0 &\cdots &\rho_{2} & 0 & 1-\rho_{3} & \cdots   & 0& 0\\
\rho_1\rho_{3}(1-\rho_{2})  & 0 & &0 & 1-\rho_{2} & 0 &  & 0& 0 \\
0 & 0 & \cdots&1-\rho_{2} & 0 & \rho_{3} &  \cdots  & 0& 0\\
 & \vdots &\iddots & \vdots&\vdots &\vdots & \ddots & \vdots& \\
0 & 1-\rho_{2} & \cdots&0 & 0 & 0& \cdots & \rho_{3} & 0\\
\CodeAfter
\end{bNiceMatrix}_{\left((2\tau+1)\times(2\tau+1)\right)},
\end{aligned}
\end{mysmall}
\end{equation}
\caption{Coefficient matrix \(\mathbf{A}\) in~(\ref{eq:idle_eq_set_example}).}
\label{fig:matrix_A}
\end{figure*}

Here,
\[
\rho_1=1-q_1,\qquad
\rho_2=1-q_2,\qquad
\rho_3=1-q_3,
\]
\[
g(i)=-\frac{1}{\tau}\big(1+(\tau-i)(1-\rho_2)\big),
\qquad
h(i)=-\frac{1}{\tau}\big(1+(\tau-i)(1-\rho_3)\big).
\]

Note that the dimension of~(\ref{eq:idle_eq_set_example}) depends on \(\tau\). We now reduce this system by merging states with equal limiting state probabilities. For the state \((\mathrm{I},\mathrm{B},k)\), where \(k\in\{0,\dots,1-\tau\}\), we have
\begin{equation}
\mathbb{I}_{(\mathrm{I},\mathrm{B},k)}=\{1\},
\qquad
\mathbb{B}_{(\mathrm{I},\mathrm{B},k)}=\{\{2\}\}.
\end{equation}
Therefore,
\[
\tilde{\pi}_{(\mathrm{I},\mathrm{B},0)}
=
\tilde{\pi}_{(\mathrm{I},\mathrm{B},-1)}
=
\cdots
=
\tilde{\pi}_{(\mathrm{I},\mathrm{B},1-\tau)}
=
\tilde{\pi}_{(\mathrm{I},\mathrm{B})}.
\]
Similarly,
\[
\tilde{\pi}_{(\mathrm{B},\mathrm{I},0)}
=
\tilde{\pi}_{(\mathrm{B},\mathrm{I},1)}
=
\cdots
=
\tilde{\pi}_{(\mathrm{B},\mathrm{I},\tau-1)}
=
\tilde{\pi}_{(\mathrm{B},\mathrm{I})}.
\]

Hence,~(\ref{eq:idle_eq_set_example}) can be compressed into
\begin{equation}
\label{eq:idle_eq_set_compressed}
\mathbf{y}^c=\mathbf{A}^c\mathbf{y}^c+\mathbf{b}^c,
\end{equation}
where
\begin{equation}
\begin{aligned}
\mathbf{y}^c
=
\left[
\tilde{\pi}_{(\mathrm{I},\mathrm{I},0)},
\tilde{\pi}_{(\mathrm{I},\mathrm{B})},
\tilde{\pi}_{(\mathrm{B},\mathrm{I})}
\right]^T,
\end{aligned}
\end{equation}
\begin{equation}
\begin{mysmall}
\mathbf{A}^c=
\begin{bmatrix}
\rho_1\rho_2\rho_3-\frac{1}{\tau} & \frac{\tau+1}{2}(\rho_2-1) & \frac{\tau+1}{2}(\rho_3-1)\\
\rho_1\rho_2(1-\rho_3) & 0 & 1-\rho_3\\
\rho_1\rho_3(1-\rho_2) & 1-\rho_2 & 0
\end{bmatrix},
\end{mysmall}
\end{equation}
and
\begin{equation}
\mathbf{b}^c=
\left[
\frac{1}{\tau},0,0
\right]^T.
\end{equation}

The reduced system~(\ref{eq:idle_eq_set_compressed}) has a dimension independent of \(\tau\), which allows the limiting state probabilities to be derived analytically as explicit functions of \(\tau\). Solving~(\ref{eq:idle_eq_set_compressed}) yields
\begin{equation}
\begin{mysmall}
\begin{aligned}
\tilde{\pi}_{(\mathrm{I},\mathrm{I},0)}
=
\frac{1}
{1 + \rho_1(\rho_3 - 1)(\rho_2 - 1)\tau^2
+ \big(1 + (-\rho_2 - \rho_3 + 1)\rho_1\big)\tau},
\end{aligned}
\end{mysmall}
\end{equation}
\begin{equation}
\begin{mysmall}
\begin{aligned}
\tilde{\pi}_{(\mathrm{I},\mathrm{B})}
=
\frac{-\rho_1(\rho_2 - 1)}
{1 + \rho_1(\rho_3 - 1)(\rho_2 - 1)\tau^2
+ \big(1 + (-\rho_2 - \rho_3 + 1)\rho_1\big)\tau},
\end{aligned}
\end{mysmall}
\end{equation}
and
\begin{equation}
\begin{mysmall}
\begin{aligned}
\tilde{\pi}_{(\mathrm{B},\mathrm{I})}
=
\frac{-\rho_1(\rho_3 - 1)}
{1 + \rho_1(\rho_3 - 1)(\rho_2 - 1)\tau^2
+ \big(1 + (-\rho_2 - \rho_3 + 1)\rho_1\big)\tau},
\end{aligned}
\end{mysmall}
\end{equation}
respectively.

\section{Transformation of~(\ref{eq:steady_embedded})}
\label{app:transformation}

This appendix explains how to transform~(\ref{eq:steady_embedded}) into a system involving only the limiting state probabilities of the states in \(\mathcal{I}\). The key idea is to eliminate the limiting state probabilities of the states in \(\mathcal{B}\) and express them in terms of those in \(\mathcal{I}\).

The transformation consists of two steps.

First, we transform the limiting state probabilities of all states in \(\mathcal{B}\) except the state
\[
(\mathrm{B}^{(1)},\mathrm{B}^{(2)},\dots,\mathrm{B}^{(N)},0,\dots,0),
\]
in which all channels are busy and start their busy periods simultaneously. This is done by using the embedded Markov chain to identify all paths from idle states to the target busy state and then accumulating the corresponding transition probabilities along these paths. In this way, each non-synchronized busy-state probability can be represented as a linear combination of the limiting state probabilities of the idle-state set.

Second, we transform the remaining limiting state probability of the fully synchronized busy state
\[
\boldsymbol{\mu}_0 = (\mathrm{B}^{(1)},\mathrm{B}^{(2)},\dots,\mathrm{B}^{(N)},0,\dots,0).
\]
To do so, we use the normalization condition
\begin{equation}
\label{eq:prob_guiyi}
\sum_{\boldsymbol{\mu}\in\mathbb{S}}
\Pr\{\text{the network state is }\boldsymbol{\mu}\text{ at time }t\}
=1.
\end{equation}
This condition states that, at any slot \(t\), the network must be in one of the states in \(\mathbb{S}\). Therefore,
\begin{equation}
\label{eq:qiuhe}
\tilde{\pi}_{\boldsymbol{\mu}_0}
=
1-
\sum_{\substack{
\boldsymbol{\mu}\in\mathbb{S}\\
\boldsymbol{\mu}\neq\boldsymbol{\mu}_0
}}
\Pr\{\text{the network state is }\boldsymbol{\mu}\text{ at time }t\}.
\end{equation}
Hence, the limiting state probability of the fully synchronized busy state can also be represented in terms of the limiting state probabilities of the idle-state set.

Combining the above two steps yields a reduced linear system involving only the limiting state probabilities of the states in \(\mathcal{I}\). This reduced system is then used in Appendix~\ref{app:limiting-probs} to derive explicit expressions for the limiting state probabilities.

\bibliographystyle{IEEEtran}
\bibliography{IEEEabrv,re}

\vfill

\end{document}